\documentclass[twocolumn,twocolappendix]{openjournal}
\usepackage{graphics}
\usepackage{amsmath}
\usepackage{amssymb}
\usepackage{hyperref}
\usepackage{color}
\usepackage[dvipsnames]{xcolor}
\usepackage{tabularx}
\usepackage{booktabs}

\hypersetup{colorlinks=true,allcolors=teal}

\newcommand{\ttimes}{{\mkern-0.25mu\times\mkern-0.25mu}}
\newcolumntype{Y}{>{\centering\arraybackslash}X}

\shorttitle{Exploring the Core-galaxy Connection }
\shortauthors{Souza Vit\'orio et al.}

\begin{document}

\title{Exploring the Core-galaxy Connection\vspace{-1.2cm }}
\author{Isabele Souza Vit\'orio\altaffilmark{3}, Michael Buehlmann\altaffilmark{2}, Eve Kovacs\altaffilmark{1}, Patricia Larsen\altaffilmark{2}, Nicholas Frontiere\altaffilmark{2}, Katrin Heitmann\altaffilmark{1}}

\affil{$^1$HEP Division, Argonne National Laboratory, 9700 S. Cass Ave.,
 Lemont, IL 60439, USA}

\affil{$^2$CPS Division, Argonne National Laboratory, 9700 S. Cass Ave.,
 Lemont, IL 60439, USA}

 \affil{$^3$Department of Physics and Leinweber Center for Theoretical
Physics, University of Michigan, 450 Church St, Ann Arbor, MI,
48103, USA}

\begin{abstract}

Halo core tracking is a novel concept designed to efficiently follow halo substructure in large simulations. We have recently developed this concept in gravity-only simulations to investigate the galaxy-halo connection in the context of empirical and semi-analytic models.  Here, we incorporate information from hydrodynamics simulations, with an emphasis on establishing a connection between cores and galaxies.  We compare cores across gravity-only, adiabatic hydrodynamics, and subgrid hydrodynamics simulations with the same initial phases.
We demonstrate that cores are stable entities whose halo-centric radial profiles match across the simulations. We further develop a methodology that uses merging and infall mass cuts to group cores in the hydrodynamics simulation, creating on average, a one-to-one match to corresponding galaxies. We apply this methodology to cores from the gravity-only simulation, thus creating a proxy for galaxies which approximate the populations from the hydrodynamics simulation. Our results pave the way to incorporate inputs from smaller-scale hydrodynamics simulations directly into large-scale gravity-only runs in a principled manner.
\end{abstract}
\keywords{methods: numerical --- cosmology: large-scale structure of the universe}

\section{Introduction}
\label{sec:intro}

Structure formation in the universe is a complex, hierarchical, multi-scale process. On the scales of the ``cosmic web'', gravity dominates and controls the formation of dark matter-dominated filaments and halos; complex astrophysical processes govern the formation of galaxies and their evolution and interaction on smaller scales within this environment. The dynamics of galaxy formation are far too complex to be treated in a fully predictive manner on cosmological scales. Various approximate empirical and semi-analytic modeling approaches exist (for reviews, see, e.g., \citealt{Mo2010}, \citealt{2018ARA&A..56..435W}), in addition to direct cosmological hydrodynamics simulations that treat galaxy formation by adding gas physics and phenomenological subgrid models for astrophysical processes, such as star and black hole formation and multiple feedback mechanisms~\citep{Somerville2015, vogelsberger2020}.

Modern cosmological sky surveys such as DESI\footnote{\url{https://www.desi.lbl.gov/for-scientists/}}, Euclid\footnote{\url{https://www.cosmos.esa.int/web/euclid}}, Rubin Observatory's LSST\footnote{\url{https://www.lsst.org/scientists/survey-design}}, the Nancy Grace Roman Space Telescope\footnote{\url{https://roman.gsfc.nasa.gov/}}, and SPHEREx\footnote{\url{https://spherex.caltech.edu/}} are characterized by large sky areas -- thousands to tens of thousands of square degrees -- and significant redshift depths ($z\sim2$ for galaxy targets and $z>6$ for quasars). To adequately model the observational results from these surveys, simulations with very large comoving volumes are needed \citep{AbacusSummit, EuclidSims,2024arXiv240607276H}. Because of the computational expense of hydrodynamics simulations -- and because of their parametric dependence on subgrid modeling assumptions -- it is often more convenient to run large-scale gravity-only simulations, within which the formation of galaxies, their properties, and other astrophysical processes (``baryonic physics'') can be modeled with detailed post-processing methods (see, e.g., \citealt{ostriker2005,Korytov2019,shaw2010,troster2019, keruzore2023,omori2024}). In this approach, galaxy formation and evolution are directly connected with the processes underlying halo formation, halo mergers, and (hierarchical) halo substructure, including subhalos, which are locally bound objects existing within larger halos.

\begin{figure}[t]
  \centering
    \includegraphics[width=\linewidth]{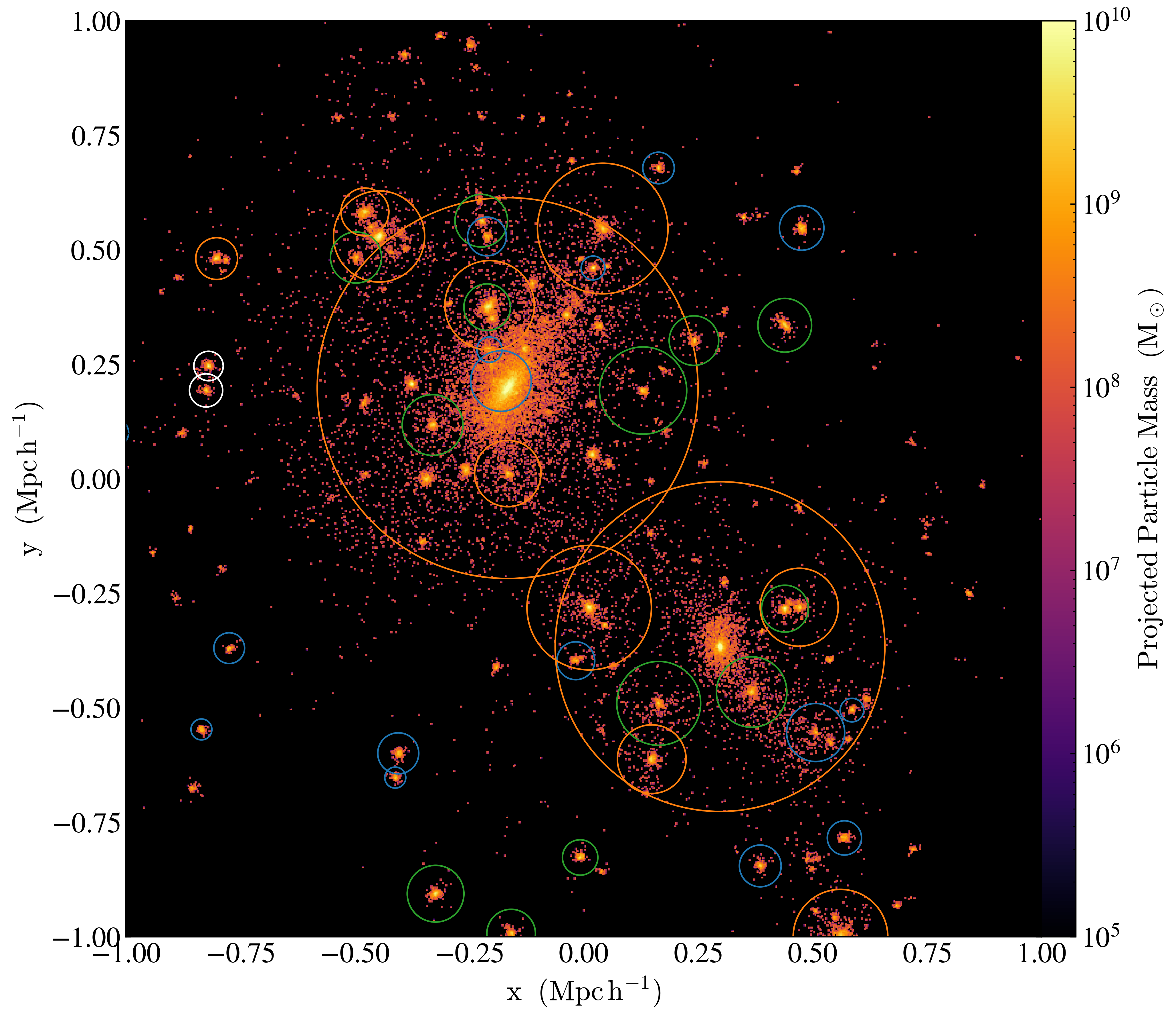}
  \caption{Visualization of a halo merger from a smaller (128 $h^{-1}$Mpc) but otherwise identical hydrodynamics simulation to the one employed in this paper. Circles indicate individual galaxies with a radius equal to the galaxy's radius. The colors of the circles indicate the number of cores associated with a galaxy, with white having no cores, blue having one, green having two, and orange having three or more.}
  \label{fig:galaxy_vis}
\end{figure}

The concept of using ``halo cores'', which are based on particle tracking methods (first used by \citealt{white1987}), was introduced as an alternative to traditional subhalo finding to establish a framework for providing a simple description of galaxy populations in halos (see \citealt{han2018, diemer2023, mansfield2023} for recent work on particle-based algorithms for identifying substructure).  A halo core corresponds to the dense inner region of a halo, which can be straightforwardly tracked in simulations~\citep{rangel2017building, korytov2023modeling}, even after the halo itself has merged into a larger structure. Halo merger trees with core tracking can be constructed efficiently~\citep{rangel2017building} and can be used as a basis for empirical models of galaxy formation and to effectively track halo substructure. Additionally, simple models for core mergers and disruption, along with simple thresholds for the halo infall mass, can be shown to provide good matches to the observed galaxy distribution in clusters as was recently shown in~\cite{korytov2023modeling}. Figure~\ref{fig:galaxy_vis} provides a visual impression of how cores and galaxies are connected. The image shows a halo merger in progress at $z=0$ with the projected particle mass density providing the background. Each circle marks an individual galaxy as identified in the simulation, with a radius equal to the galaxy's radius. 

The physical motivation for using cores is as follows: first, as in subhalo abundance matching (SHAM; \citealt{Kravtsov2004}, \citealt{Vale2004}, \citealt{Conroy2006}), we assume that all galaxies form as ``central'' galaxies in halos and become satellites only through subsequent merging with other halos. Second, we note that the parent halo's central star-rich high-density region is resistant to disruption and that stars follow gravity in the same way as the dark matter, so gas dynamics may not be important as far as the cores are concerned, at least as a first order approximation. The validity of this assumption will be further investigated and established in this paper. 

The choice to use cores over subhalos for modeling the galaxy-halo connection is motivated by several well-known deficiencies of subhalos as galaxy proxies, most notably when looking at group and cluster-scale halos \citep{vdBosch2017, vdBosch2018a, vdBosch2018b} or at simulations with low mass resolution \citep{errani2021}. Within large halos, subhalos can readily lose mass inside the halo's virial radius due to the effects of gravitational tidal forces (\citealt{Wetzel2010}) -- while the corresponding galaxies, being much more compact in their stellar radius, are much less impacted. Additionally the effect of insufficient mass resolution in the simulation~(\citealt{Guo2014}, \citealt{vdBosch2018a}, \citealt{errani2021}) can make it impossible to capture smaller substructures using subhalo finding, and the (post-accretion) subhalo mass by itself is an unstable quantity (being subject to tidal forces and mergers), and so may not accurately predict the properties of the galaxies hosted within these subhalos. This is particularly true for properties that are sensitive to particles at the outer edge of subhalos \citep{onions2012}. We stress that our intention is not to use cores as a substitute or proxy for subhalos.  We have argued above that, compared to subhalos, cores are a more reliable and stable proxy for {\it galaxies}, and therefore have the potential to provide a more robust galaxy-halo connection model. 

When addressing the longer term goal of establishing an accurate galaxy-halo connection, the galaxy locations and velocities are only part of the picture. 
It is equally important to correctly describe the many properties of the simulated galaxies (such as the luminosity, stellar mass, color, and morphology). Within the landscape of current semi-analytic and empirical models for galaxy formation, cores can function as an essential component in providing this information. When required, a subhalo-like mass can be associated with cores, for example by using the SMACC (Subhalo Mass-loss Analysis Using Core Catalogs) model described in~\cite{sultan2021last}, which uses the mass loss model of~\cite{2005MNRAS.359.1029V} applied to the infall halo mass associated with a core, validated against a separate subhalo finder. This additional mass assignment ability enables the use of cores within existing technologies such as semi-analytic models for galaxy formation to provide a complete picture of the galaxy; in this case they can also help to address the problem of ``orphan galaxies'' (those not associated with subhalos) as discussed in \cite{campbell2018, derose2022}.

In this paper we seek to investigate how cores can contribute to more effective galaxy modeling, by both studying cores across gravity-only and hydrodynamics simulations, and comparing these cores to their galaxy counterparts. We investigate key elements to this modeling approach, including core merging and infall into the halo center (or central galaxy). Similar elements have been investigated previously using simple models for core disruption and merging, and the results compared with observations of SDSS cluster galaxies in~\cite{korytov2023modeling}. However, the availability of hydrodynamics simulations now allows for a much more controlled study.

To summarize, our aims in this paper are threefold: we 1) explore the effects of baryonic physics on core stability by comparing gravity-only, nonradiative or adiabatic hydrodynamics, and ``full'' hydrodynamics simulations (including cooling and subgrid models for galaxy formation and feedback), 2) compare core distributions across hydrodynamics and gravity-only simulations in a controlled setting, and 3) investigate the core-galaxy connection within a hydrodynamics simulation with a specified set of subgrid modeling parameters related to galaxy formation. To address these aims, we employ three simulations (gravity-only, adiabatic, and full hydrodynamics), all started from matched initial conditions. 

The results of our investigations establish that cores are stable entities
with very similar large-scale structures 
across the simulations with different physics effects included. 
Furthermore, cores have radial distributions that are 
tightly correlated across the simulations, and have a direct (and simple) relationship to the associated galaxy objects in the hydrodynamics simulation. 

The paper is organized as follows. In Sec.~\ref{sec:sims} we provide a description of the gravity-only and hydrodynamics simulations. In Sec.~\ref{sec:stats} we investigate and compare core properties across the three different simulations. 
In Sec.~\ref{sec:coresandgalaxies} we study the galaxy-core connection. We provide conclusions and an outlook in Sec.~\ref{sec:conclusion}, where we discuss further applications of the work presented here.

\section{Simulations}
\label{sec:sims}
In this paper, we employ a set of gravity-only and hydrodynamics simulations carried out with the Hardware/Hybrid Accelerated Cosmology Code (HACC) framework~(see \citealt{Habib:2014uxa} for a description of the gravity-only code and~\citealt{frontiere_crkhacc} for the hydrodynamics-enhanced version). The simulations use as input the best-fit Planck cosmology~\citep{Planck_params} specified by $\Omega_{\rm cdm}=0.26067, \omega_b=0.02242, h=0.6766, \sigma_8=0.8102, n_s=0.9665, w=-1$, assuming a flat universe and massless neutrinos. All simulations are initialized at $z=200$ using the Zel'dovich approximation~\citep{zel1970gravitational} and are evolved forward to $z=0$ using the same number of global time steps and the same initial conditions with the same random seed to enable a direct comparison of the results. Each run covers a volume of $V=$~(576\,$h^{-1}$Mpc)$^3$. We initialize all species using the total matter transfer function computed with CAMB~\citep{2000ApJ...538..473L}. The gravity-only simulation is run with 2304$^3$ particles and the hydrodynamics simulations evolve twice as many to account for the dark matter and baryonic particle species separately. The nominal fixed comoving force resolution (Plummer force softening) of 10\,$h^{-1}$kpc is the same for all simulations. A summary of the important simulation parameters is provided in Tab.~\ref{tab:sims}.

\begin{table}[t]
  \begin{center}
  \caption{Simulation Parameters\label{tab:sims}}
    \begin{tabularx}{\linewidth}{@{}X|rrrr@{}}
      Type & Particles & $m_{m}$  & $m_{b}$ & $m_{\rm cdm}$ \\
      \midrule
      Gravity-only     & $2304^3$ & $1.34\ttimes 10^9$ & n/a & n/a\\
      Adiabatic        & $2 \ttimes 2304^3$ & n/a   & $2.12 \ttimes 10^8$ & $1.13\ttimes 10^9$\\
      Full (w/ subgrid) & $2 \ttimes 2304^3$ & n/a  & $2.12 \ttimes 10^8$ & $1.13 \ttimes 10^9$\\
   \end{tabularx}
  \end{center}
  {\sc Tab.~\ref{tab:sims}.---} Particle numbers and masses for the simulations used in this work. The linear box size is 576\,$h^{-1}$Mpc for all of the simulations; the initial conditions have the same phases. The particle mass units are $h^{-1}$M$_\odot$. See the text for the subgrid model descriptions used in  the full hydrodynamics simulation.
\end{table}

\subsection{Gravity-only Simulations}
\label{sec:go}
The gravity-only simulation is carried out with HACC's spectral P$^3$M GPU solver~\citep{Habib:2014uxa} on Summit, a GPU-accelerated supercomputer at the Oak Ridge Leadership Computing Facility. A range of analysis tools are run on-the-fly with the simulation, including a halo finder, power spectrum measurements, light-cone generation, and halo core tracking at 101 snapshots, evenly spaced in log$_{10}a$ ($a$ is the scale factor). For our current purposes, the key analysis tools are the halo finder and the core-tracking infrastructure.

As part of our analysis, we run a Friends-of-Friends (FoF) halo finder on the non-collisional particles at 101 timesteps, with a linking length of $b_{\rm FoF}=0.168$\footnote{Here, the linking length $b_{\rm FoF}$ is given in units of the interparticle spacing in the simulation.}, starting at $z\sim10$.  We record a range of halo properties (for a detailed list see, e.g., \citealt{2021ApJS..252...19H}) for halos with at least 20 particles. 
For this analysis, we have used SO halo masses which rely on the FoF algorithm only to identify the halo center of halos with at least 100 particles. Once identified, we build corresponding SO halos around the potential minimum, out to the corresponding radii used in the analysis.
Section~\ref{sec:cores} presents more details about the process to extract and organize halo core information.

\subsection{Hydrodynamics Simulations}
\label{sec:hydro}

The cosmological hydrodynamics simulations are conducted with CRK-HACC~\citep{frontiere_crkhacc}, which uses an improved smoothed particle hydrodynamics (SPH) solver, CRKSPH (Conservative Reproducing Kernel SPH, \citealt{frontierecrk}). As stated above, two hydrodynamics simulations are carried out, a non-radiative run (``adiabatic'') and one that includes a full suite of astrophysical subgrid models. The adiabatic hydrodynamics simulation includes the effects of gas dynamics (baryonic pressure forces), but no additional astrophysical modeling. While the details of the subgrid model packages and calibration efforts will be described elsewhere, in the following, we briefly summarize the models implemented and calibration metrics and parameter values utilized in this work. 

\begin{figure}
  \centering
  \includegraphics[width=\linewidth]{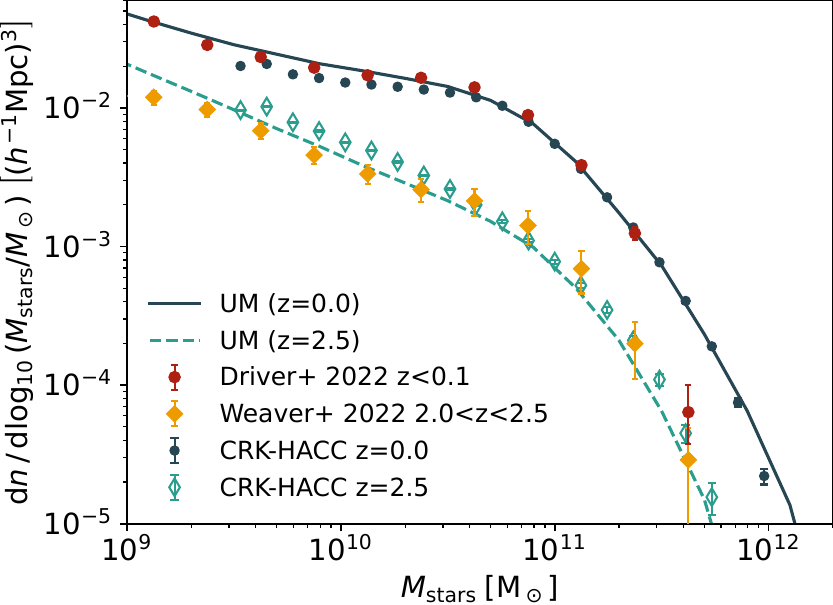}
  \caption{Galaxy stellar mass function from the hydrodynamics (w/ subgrid) CRK-HACC simulation at redshifts $z=0$ and $z=2.5$. For comparison, the stellar mass function constraints provided by the Universe Machine (UM) model \citep{behroozi2019universemachine} 
  and observational results from the GAMA survey \citep{Driver2022} and the COSMOS survey \citep{weaver2022, weaver2023} are shown.}
  \label{fig:GSMF}
\end{figure}

The additional physics modules in CRK-HACC include 1) a radiative cooling and heating model that assumes gas to be optically thin and present in a spatially uniform time-varying ultraviolet background described by~\cite{faucher2020}, 2) metal-line cooling treatment similar to~\cite{wiersma2009chemical}, with cooling rates that are calculated using the simulation code CLOUDY\footnote{\url{https://gitlab.nublado.org/cloudy/cloudy/-/wikis/home}}, 3) star formation and supernova feedback that are incorporated using a modified hybrid multi-phase model originally described in~\cite{springel2003cosmological}, in addition to a chemical enrichment model which integrates enrichment rates from the FIRE simulations \citep{hopkins2018fire}, 4) a kinetic galactic outflow (wind) model implemented similar to the TNG prescription (\citealt{pillepich2018} with wind velocity and energy parameter values $\kappa_w = 3.7$ and $e_w=0.5$), and 5) an active galactic nuclei (AGN) model with  details also closely following TNG, where we take the kinetic feedback model from~\cite{weinberger2016simulating} (using matching parameter values for both the momentum feedback and $\chi$ thresholds), and tune the black-hole seed mass ($m_\text{BH,seed}$) and black-hole interaction neighbor count ($n_\text{BH}$) to our calibration targets defined below. 
In the simulations presented here, $m_\text{BH,seed} = 1.25 \times 10^6$\,$h^{-1}\text{M}_\odot$ and $n_\text{BH} = 32$ interacting gas neighbor extent. 

\begin{table*}[t]
  \begin{center}
  \caption{Central and satellite core counts across simulations  \label{tab:cores_halo}}
  \small
 \begin{tabularx}{\linewidth}{@{}Xcccccccc@{}}
    \multicolumn{1}{c}{} & \multicolumn{4}{c}{Central Cores} & \multicolumn{4}{c}{Satellite Cores} \\
    \cmidrule(r){2-5} \cmidrule(l){6-9}
    \raggedright $\log_{10}$(M$_h/(h^{-1}$M$_\odot))$  & $[11.3, 12]$ & $[12, 13]$ & $[13, 14]$ & $\geq 14$  & $[11.3, 12]$ & $[12, 13]$ & $[13, 14]$ & $\geq 14$ \\
    \midrule

    Gravity-only        & $2.02 \times10^6$ &  $5.71\times10^5$ & $6.32\times10^4$ & $3.53\times10^3$ &  $1.47\times10^5$ & $1.05 \times10^6$ & $1.10\times10^6$ & $4.15\times10^5$ \\
    Adiabatic           & $1.83\times10^6$ & $5.62\times10^5$   & $6.44\times10^4$ & $3.58\times10^3$ &   $1.30\times10^5$& $9.93\times10^5$ & $1.06\times10^6$ & $3.92\times10^5$\\
    Full (w/ subgrid)  & $2.30\times10^6$  & $5.48\times10^5$  & $6.16\times10^4$ & $3.61\times10^3$  & $2.18\times10^5$ & $1.27\times10^6$ & $1.27\times10^6$ & $4.78\times10^5$\\
  \end{tabularx}
  \end{center}
{\sc Tab.~\ref{tab:cores_halo}.---} Central and satellite core counts for four SO halo mass bins for each simulation used in this work. The number of central cores is by definition equal to the number of halos in each halo mass bin in a given simulation; the halo mass units are $h^{-1}$M$_\odot$. The number of satellite cores per halo is larger than the typical number of subhalos that we would expect to find, since the possibility of merging cores has not been taken into account. We will return to this point in Sec.~\ref{sec:coresandgalaxies}. 
The highest mass bin includes all halos with $M_{200c} \geq 10^{14}h^{-1}$M$_{\odot}$. Overall, the number of central cores is very similar across the three simulations. The number of satellite cores is also similar, with the exception of those for the lower mass bins in the full hydrodynamics simulation, which are $\sim$ 50\% higher than the number for the other simulations. 
\end{table*}

The values of the four primary active model parameters ($\kappa_w$, $e_w$, $m_\text{BH,Seed}$, and $n_\text{BH}$) described above are calibrated to match observations, with the galaxy stellar mass function (GSMF) as a main constraint. Figure~\ref{fig:GSMF} shows the calibrated GSMF for the subgrid simulation used in this work, compared to observational data. In our measurements, the galaxy stellar mass is defined as the mass of star particles contained in an annulus of radius 30 kpc in proper units around a galaxy center (as was done similarly in, e.g., \citealt{schaye2015eagle}). Furthermore, the CRK-HACC galaxy finder employs the publicly available ArborX code\footnote{\url{https://github.com/arborx/ArborX}} \citep{arborx2020} to efficiently find galaxies using a GPU-accelerated density-based spatial clustering (DBSCAN) algorithm applied to the star particles. DBSCAN clusters are generalized FoF objects, which not only enforce a minimal connecting linking length ($b_\text{DB}$), but also require a minimum number of linking neighbors ($n_\text{DB}$) per cluster particle member (where $n_\text{DB} = 1$, for the standard FoF algorithm). The simulation is run with $b_\text{DB} = 30$ kpc in proper units and $n_\text{DB} = 10$, i.e. galaxies are found as clusters of connected star particles (with 10 neighbors), where the stellar potential center is then found and utilized as the location of the 30 kpc annuli cutout measurements for the galaxy stellar mass.

We note that, while the GSMF in the hydrodynamics run is a reasonably good match to observations, as shown in Fig.~\ref{fig:GSMF}, it is not our purpose here to discuss how faithfully a wide set of galaxy properties can be reproduced in such a simulation. Our main aim, instead, is to establish that the spatial distribution of galaxies as predicted by the hydrodynamics simulation can be (approximately) mirrored in a gravity-only simulation. As the hydrodynamics simulations improve in fidelity, the properties of the simulated galaxies will come closer to reality, but the spatial matching process will remain largely unchanged. For the rest of the paper, given the stellar mass function results, we further impose a stellar mass cut on the galaxy catalog of 100 stellar particles which corresponds to a minimum galaxy aperture stellar mass of 
$1.87 \times 10^{10}$\,$h^{-1}$ M$_\odot$, which is well into the region of good agreement between the simulated GSMF and the observations shown in Fig.~\ref{fig:GSMF}. This agreement provides good evidence that 100 star-particle clusters identified by DBSCAN are indeed stable galaxies and are not merely spurious associations of star particles. Furthermore, we examined the correlation between the DBSCAN masses and the 30 kpc aperture masses and found good consistency between the two determinations. The DBSCAN masses are higher than the aperture masses, as expected, since the aperture radius is equal to $b_\text{DB}$.

\subsection{Core Merger Trees}
\label{sec:cores}
As discussed in Sec.~\ref{sec:intro}, core tracking offers a method for following substructure within halos with some advantages over the more traditional use of subhalos. The core approach relies on tracking particles closest to halo centers, avoiding the computationally more expensive subhalo finding algorithms. 
In this section, we detail our methodology for measuring and tracking cores, including the information of halo mergers to provide a comprehensive substructure measurement with merging histories.  

The definition of halo cores and the use of cores to track substructures within the HACC framework was first introduced in~\cite{rangel2017building} and further described in~\cite{2021ApJS..252...19H}. In~\cite{sultan2021last}, core merger trees were used to establish the connection with subhalo descriptions and \cite{korytov2023modeling} employed cores to predict the distribution of galaxies in cluster-sized halos, with comparisons to data from the Sloan Digital Sky Survey (SDSS). 

The first step to identify cores involves running the FoF halo finder on-the-fly at each analysis time snapshot. For each FoF halo found with at least 80 particles, we mark the 50 particles closest to the halo center as core particles and store their particle ids, positions, velocities and halo ids.\footnote{Particle ids are uniquely assigned at the start of the simulation and never change. Halo ids are re-assigned at every analysis time snapshot and we track the history of halos in post-processing via the construction of halo merger trees.}  Once a particle is identified as a core member, we continue tracking it at subsequent analysis steps, storing its updated position.

In post-processing, we follow the procedure described in~\cite{rangel2017building} for building halo merger trees. The tree construction is based on matching particles in halos across time steps using their unique ids and recording all progenitors. Halo evolution includes halo mergers and splitting which are both taken into account. After the halo merger trees have been constructed, we incorporate the core information to generate core merger trees. A core is designated as a ``central'' if it is the core that was associated with the most massive progenitor of the halo and core particles are updated based on consistency between the current central core and the central core from the previous timestep. 
In cases of discrepancy, we choose to track existing cores rather than update with newly measured core particles to avoid unphysical evolution. Since we track all core particles throughout time, we can identify cores within halos and monitor their evolution after infall into new host halos. Since the vast majority of satellite galaxies originate from infallen halos, such cores serve as proxies for the positions and properties of satellite galaxies, though they may not exactly match those from simulations with full baryonic physics. In the context of this paper, core merger trees are primarily used to determine halo infall times and to map cores to their host halos.


\section{Cores across Simulations}
\label{sec:stats}
In this section, we compare core properties across the different simulations to investigate the impact of gas physics and subgrid models on the cores, treating the gravity-only simulation as a reference. We use all three simulations (gravity-only, adiabatic hydrodynamics, and hydrodynamics with subgrid models) and measure a range of core statistics. We consider only cores found in halos with a minimum SO halo mass of M$_{200c}=2\times 10^{11}$\,$h^{-1}$M$_\odot$, corresponding to $\sim$150 simulation particles in the gravity-only simulation. Note that while we built the core merger trees from FoF halos, for the halo mass bins in our analysis, we choose to use SO masses for our comparisons since, in the hydrodynamics simulations, SO masses include contributions from all particle species. Furthermore, the centers of the SO halos are the same as the centers for the FoF halos by definition, making the FoF core statistics equivalent to SO core statistics. 

The comparisons across the three simulations address the first aim of exploring the effects of baryonic physics on core properties. If the baryonic effects are small, or at least can be modeled sufficiently well, then, if necessary, smaller volume hydrodynamics simulations can be used to develop baryonic corrections that can be applied to much larger gravity-only runs. In the following, we compare core counts, core sizes, and the radial distribution of cores in halos and illustrate a visual comparison of the spatial distribution of cores across the three simulations. 

\subsection{Core Counts}
\label{subsec:core_counts}
We first compare the total number of cores 
at $z=0$ in four halo mass bins between the gravity-only and the two hydrodynamics simulations.
Results are shown in Tab.~\ref{tab:cores_halo}, divided into central and satellite cores. The introduction of baryonic effects alters the central core counts in the three higher mass bins by less than 5\% relative to the gravity-only simulation. 
For the lowest mass bin, for halos with masses between 2$\times 10^{11}$\,$h^{-1}$M$_\odot\leq$ M$_{200c}$ $<10^{12}$\,$h^{-1}$M$_\odot$, the central core counts differ by 10-15\%. In general, the adiabatic hydrodynamics simulation has fewer central cores compared to the gravity-only simulation due to the baryonic pressure forces that push the baryons out of the halo and therefore lead to smaller halo masses. Since cores are identified only in halos above a mass threshold, this has the effect of reducing the number of cores in the adiabatic hydrodynamics simulation. The hydrodynamics simulation with subgrid modeling shows the opposite trend, with more central cores compared to the gravity-only simulation. We surmise that this is due to gas cooling, which produces more compact local density enhancements, creating more halos and therefore cores. 

Compared to the central cores, the differences in counts of satellite cores are relatively higher. One reason is that
the number counts across simulations for central cores differ more for the lower mass bins which are the source for most satellite cores.
Therefore, we expect the differences in the counts for satellite cores to be amplified across all mass bins compared to the counts for central cores.
Comparing the gravity-only and adiabatic hydrodynamics simulations, the three higher mass bins show a reduction in counts of around 5\%, and for the lowest mass bin a reduction of $\sim$10\%. The increase in satellite core counts after introducing subgrid modeling are significantly larger: $\sim$15\% for the three higher mass bins and 50\% for the lowest mass bin. Note that the physics underlying these changes in core counts is similar to that of the central cores, as each satellite core previously resided in an isolated halo in the structure formation process. 

In addition to comparing the absolute satellite core counts, we also briefly contrast the number of satellite cores per halo across the simulations. This is an important comparison for our main goal to use cores as tracers of substructures and eventually the galaxies within halos. Using the satellite core counts in Tab.~\ref{tab:cores_halo}, we find that for the lowest mass bin,
there are on average 0.07 satellite cores per halo in the gravity-only and adiabatic hydrodynamics simulations, compared to 0.1 in the subgrid  hydrodynamics simulation. In the second lowest mass bin, the first two simulations are again very close, with on average 1.8 satellite cores per halo, as compared to 2.3 in the full hydrodynamics simulation. For the second highest mass bin, the average number of satellite cores per halos across the simulations is 17, 16, and 20 in the gravity-only, adiabatic and full hydrodynamics simulations respectively. And finally for the cluster-size halos above a mass of M$_{200c}=10^{14}$\,$h^{-1}$M$_\odot$, the average number of satellite cores per halo is 118, 109, and 132 for the three simulations, respectively. Overall, the results are close especially between the gravity-only and adiabatic hydrodynamics simulations, with values from the subgrid hydrodynamics simulation showing a modest 10\% to 20\% excess above the values for the other two simulations.

\begin{figure}[b]
  \centering
  \includegraphics[width=\linewidth]{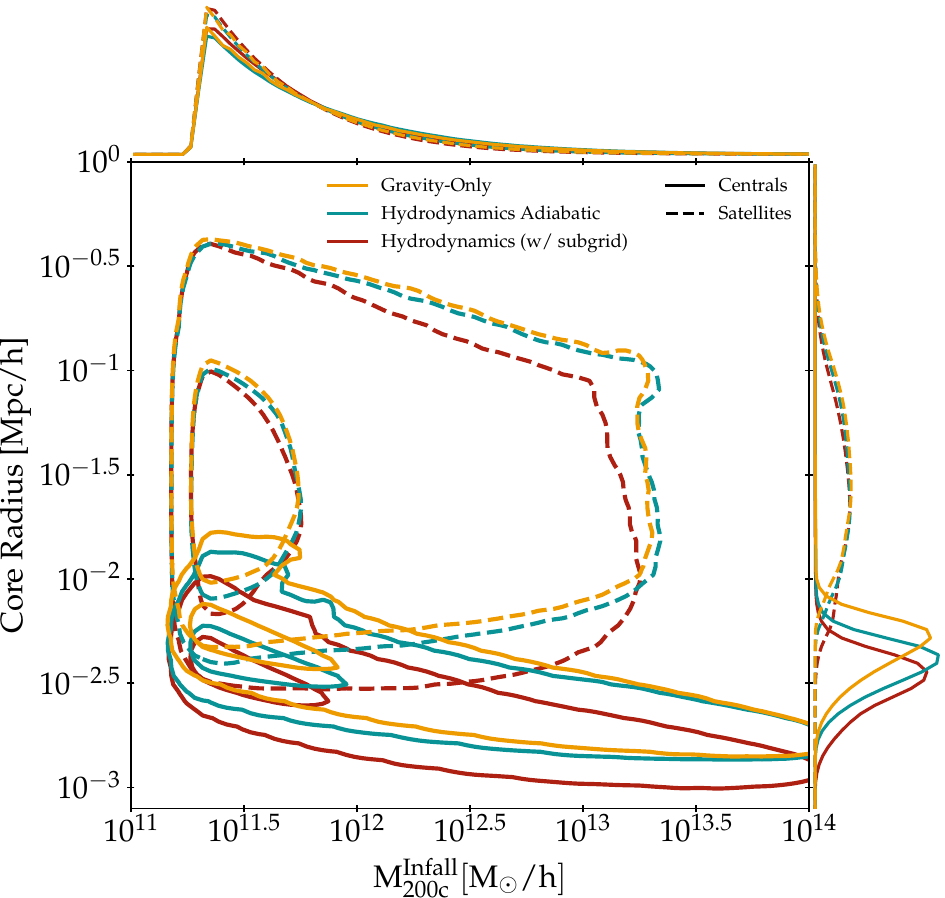}
  \caption{Distributions of core radii and infall SO halo masses across different simulations at z=0. Shown are the individual one-dimensional distributions (top and right) and the combined two-dimensional distribution (main panel), where the inner and outer contours enclose 10\% and 97\% of the cores respectively.
  For central cores (solid lines), we use the host SO halo mass at $z=0$, whereas for satellite cores (dashed lines) we use the SO halo mass at the redshift of that halo's first infall. A host halo mass cut of $2\times 10^{11}$\,$h^{-1}$M$_\odot$ is applied to all the cores in this figure. Central cores generally have smaller radii than satellite cores. The overlap between the two distributions arises from undisrupted satellite cores that have either recently merged or that reside in the central regions of their host halos.}
  \label{fig:core_rad}
\end{figure}

\begin{figure*}[t]
  \centering
  \includegraphics[width=5.5in]{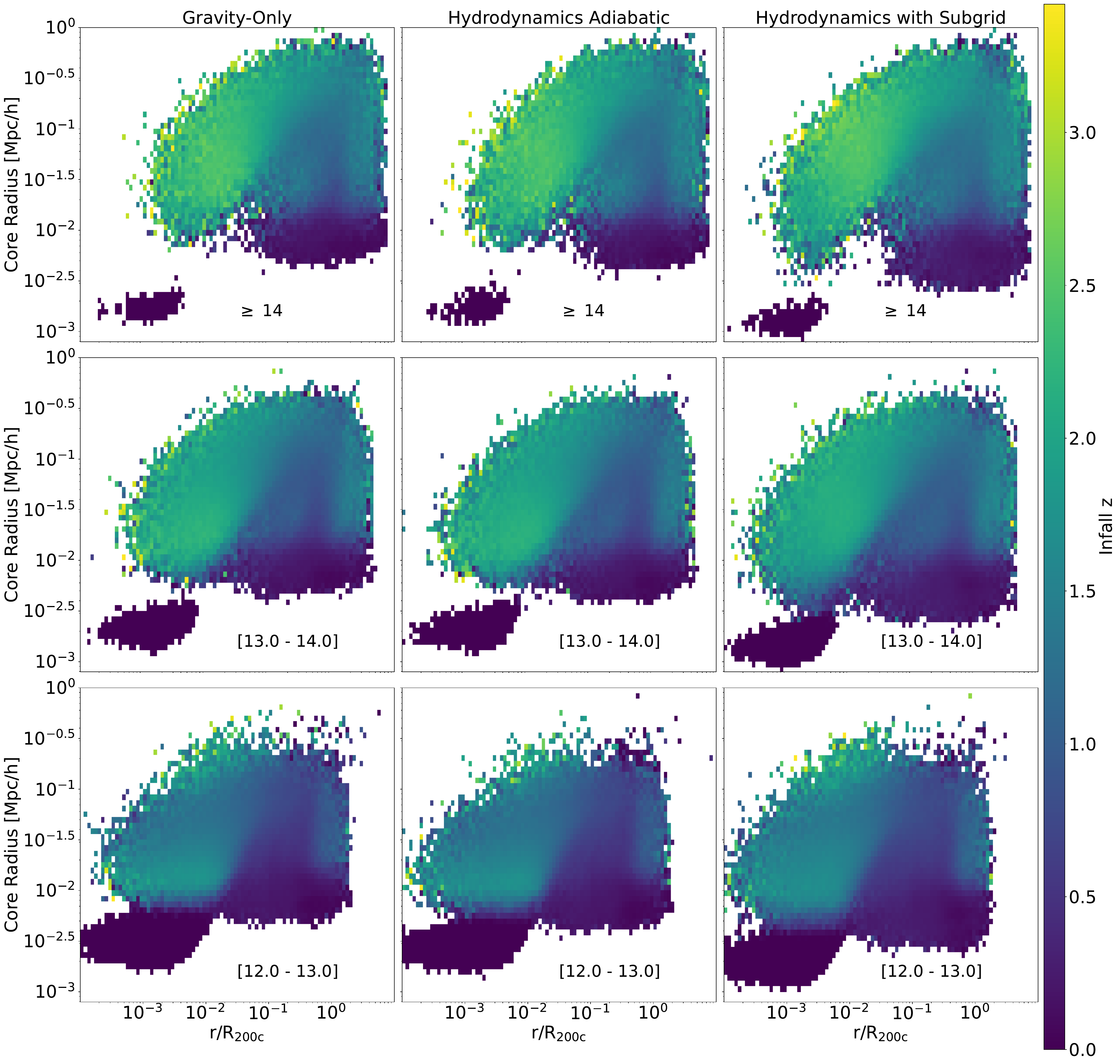}
  \caption{Distributions of the core radius versus $r/R_{200c}$ at $z = 0$ across the simulations for the three highest mass bins from Tab.~\ref{tab:cores_halo}. The colorbar shows the average infall redshift for each bin. We show only satellite cores with infall redshifts less than $z=3.5$. In each plot, the small dark blue cloud with small values of the core radius and $r/R_{200c}$ is comprised of central cores. The broad distribution with a range of $r/R_{200c}$,  $R_{\rm{eff}}$ and infall redshift values is comprised of three subpopulations of satellite cores (see text for more details). Overall, the distributions are similar across simulations and mass bins. The most notable trend is the tightening of the central core radii with increasing halo mass.}
  \label{fig:CoreRadius_dist2}
\end{figure*}

\subsection{Core Sizes and Halo Masses}
In this section, we investigate the spatial extent of individual cores. The core size can be an indicator of disruption within the host halo, with a large core radius signaling that a previously compact object has been subject to tidal forces that have ripped it apart. Conversely, the core size can remain small for satellite cores that have fallen into the center of their host halo but can nevertheless be considered as disrupted due to a potential  merger with the central core of the halo. In~\citealt{korytov2023modeling}, where cores were used to model the radial profiles of SDSS cluster galaxies, the core radius was one of the modeling parameters (along with the infall mass and a core merging parameter) employed to tune the model to agree with the observations.  Hence, it is important to understand the consistency of the size distributions across the simulations. 

Following~\cite{korytov2023modeling}, we evaluate the relationship between the average core radius and the SO halo mass of the associated halo. The core radius, $R_{\rm{eff}}$, is defined as the {\em rms} standard deviation of the core particle positions relative to the average position as in \cite{korytov2023modeling}:
\begin{equation}
    R_{\rm{eff}}=\frac{1}{3}\sqrt{\frac{1}{N}\sum_{i=1}^N (\Delta x_i^2+\Delta y_i^2+\Delta z_i^2)},
\end{equation}
where $N$ is the number of core particles and the coordinate distances are evaluated over the individual core particles relative to the average position of all the core particles. 
For central cores, the SO halo mass of the associated halo is the host halo mass, while for satellite cores, it is their halo mass at the time of first infall into a host halo. The results for the two-dimensional distributions of core counts as a function of core radius and SO halo mass are shown in Fig.~\ref{fig:core_rad}. The one-dimensional distributions of core counts versus SO halo mass and core radius are shown at the top and right-hand side of the figure, respectively. As expected, there are two classes of cores: 1) the more compact halo-centric cores (solid lines) and  2) a more dispersed population of satellite cores (dashed lines) that is subject to tidal forces during their migration towards the center of the halo. These tidal forces can lead to an increase in core size and eventually complete disruption. The two size distributions overlap as 1) recently merged cores will have smaller levels of disruption, with sizes similar to that of central cores, and 2) satellite cores that fall into the central potential well can also be compact. Figure~\ref{fig:core_rad} also shows that the distribution of core sizes is relatively insensitive to baryonic physics, with the adiabatic (light blue) and gravity-only (orange) distributions being very close, while the full hydrodynamics simulation (red) has a tendency towards smaller core sizes, likely due to the previously mentioned effects of baryonic cooling.

Figure~\ref{fig:CoreRadius_dist2} shows two-dimensional histograms of the core radius as a function of the halo-centric distance for three different mass bins.  The colors depict the average infall redshift of the cores, with the infall redshift of central cores being set to the current redshift, $z = 0$. The infall redshift for each satellite core is determined by the redshift of the first infall of the core's initial host halo into another halo and thus corresponds to the redshift at which the core transitioned from being a central to a satellite core.

The patterns exhibited by these distributions are quite similar across the three simulations. Qualitatively, they can be characterized by the same general features, consisting of a small cloud with infall redshift $z=0$ and small values of the core radius and $r/R_{200c}$, and a much broader cloud with a range of core radii, halo-centric distances and infall redshifts. As for Fig~\ref{fig:core_rad}, the small dark blue cloud corresponds to compact cores close to the halo center, while the large multicolored cloud corresponds to satellite cores at various stages of infall to the center of their host halo. The most striking feature of the figure is the strong similarity of the color patterns across all three simulations. This means that radial distributions of cores as a function of their infall time are very similar, despite the additional baryonic effects included in the hydrodynamical simulations. We explore this idea further in Sec.~\ref{subsec:CoreMatching}.

\subsection{Radial Core Counts}
Next, we study the spatial distribution of cores as a function of halo-centric radius.  We first evaluate if the core distributions in the three simulations are similar; later, in Sec.~\ref{sec:coresandgalaxies}, we will compare the distribution of cores against that of galaxies from the full hydrodynamics simulation to further elucidate the core-galaxy connection.

\begin{figure}[t]
  \centering
    \includegraphics[width=3in]{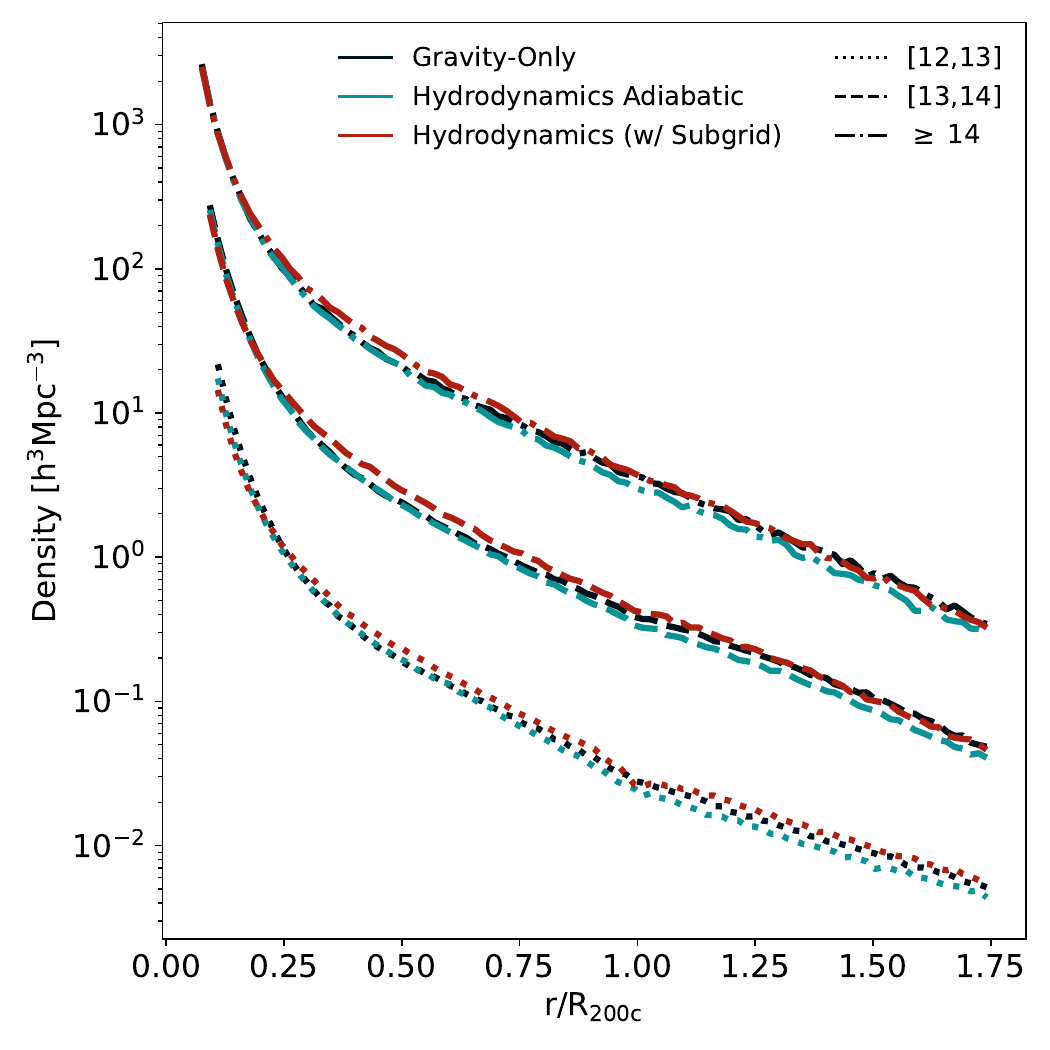}
  \caption{Core density as a function of $r/R_{200c}$ at $z = 0$ across the simulations for the three highest SO halo mass bins from Tab.~\ref{tab:cores_halo}. As described in the text, we consider only values of $r/R_{200c}$ above a threshold set by the force resolution of the simulation and the average value of $R_{200c}$ for each halo mass bin. The core densities are very similar across the three simulations.}
  \label{fig:core_count_beforecut}
\end{figure}

Figure~\ref{fig:core_count_beforecut} presents the normalized core density as a function of the normalized distance from the halo center, $r/R_{200c}$, at $z=0$ across the simulations for the three highest SO halo mass bins listed in Tab.~\ref{tab:cores_halo}. For each bin in $r/R_{200c}$,  we calculate the average core count per halo and normalize by the volume of the spherical shell enclosed by each bin. Since objects in the simulation are resolved to a few times the force softening scale, the profiles are not reliable below this scale. We therefore show only values of $r/R_{200c}$ that exceed $0.046$, $0.021$, $0.010$ for the lowest to highest mass bins, respectively, where these threshold values are obtained by dividing the nominal force resolution of the simulations (10\,$h^{-1}$kpc) by the average value of $R_{200c}$ for each mass bin.  We note that all three simulation results yield very similar profiles with the following minor differences: 1) within the SO halo radius, ($r/R_{200c} \lesssim 1$) but outside of the central region ($r/R_{200c} > 0.2$), the density profile for the subgrid hydrodynamics simulation exceeds those of the other two simulations, which are very close to each other. 2) Outside of the SO halo radius ($r/R_{200c} \gtrsim 1$), the profiles for the subgrid hydrodynamics and gravity-only simulations are very close and above the adiabatic hydrodynamics simulation. 3) Finally for the inner central region ($r/R_{200c} \lesssim 0.2$), all profiles for the two highest mass bins are very close, whereas the profile for the gravity-only simulation lies slightly above the others for the lowest mass bin. These differences are consistent with the expected trends due to the different physics effects included in the simulations and discussed in Sec.~\ref{subsec:core_counts}: gas pressure effects tend to reduce the core profiles for the adiabatic hydrodynamics simulation, while gas cooling effects favor the increase of core profiles in the subgrid hydrodynamics simulation, but subgrid feedback effects act to reduce the profiles at the centers of halos. 
The overall similarity in the profiles means that if we find a prescription for identifying cores with galaxies within the full hydrodynamics simulation, we can expect this prescription to apply to cores in the gravity-only simulation. Hence, we can reproduce the expected radial distribution of galaxies by using the gravity-only cores as galaxy tracers. We will return to this topic in Sec.~\ref{sec:coresandgalaxies}.

\subsection{Spatial Distribution of Cores and Halos}

\label{subsec:CoreMatching}
\begin{figure*}[t]
  \begin{center}
    \includegraphics[width=0.29\textwidth]{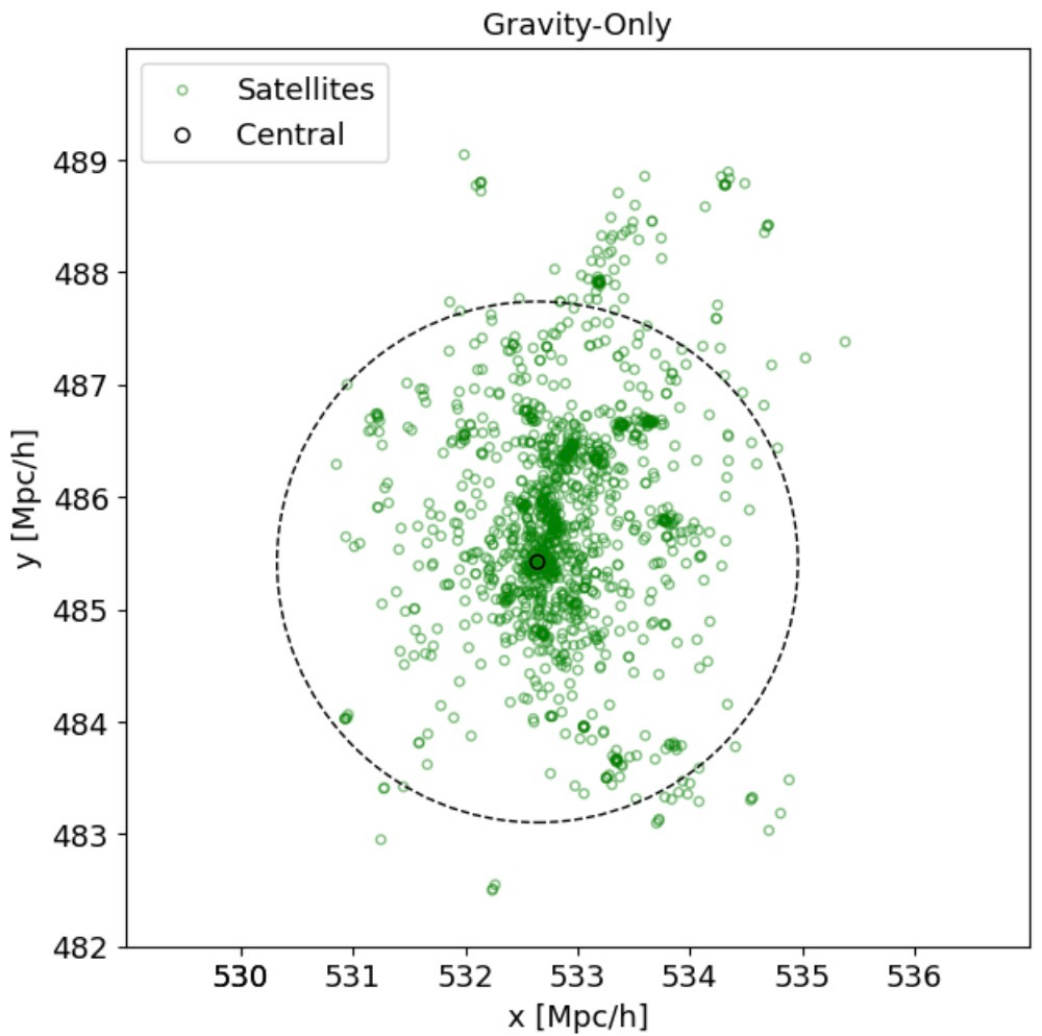}\includegraphics[width=0.29\textwidth]{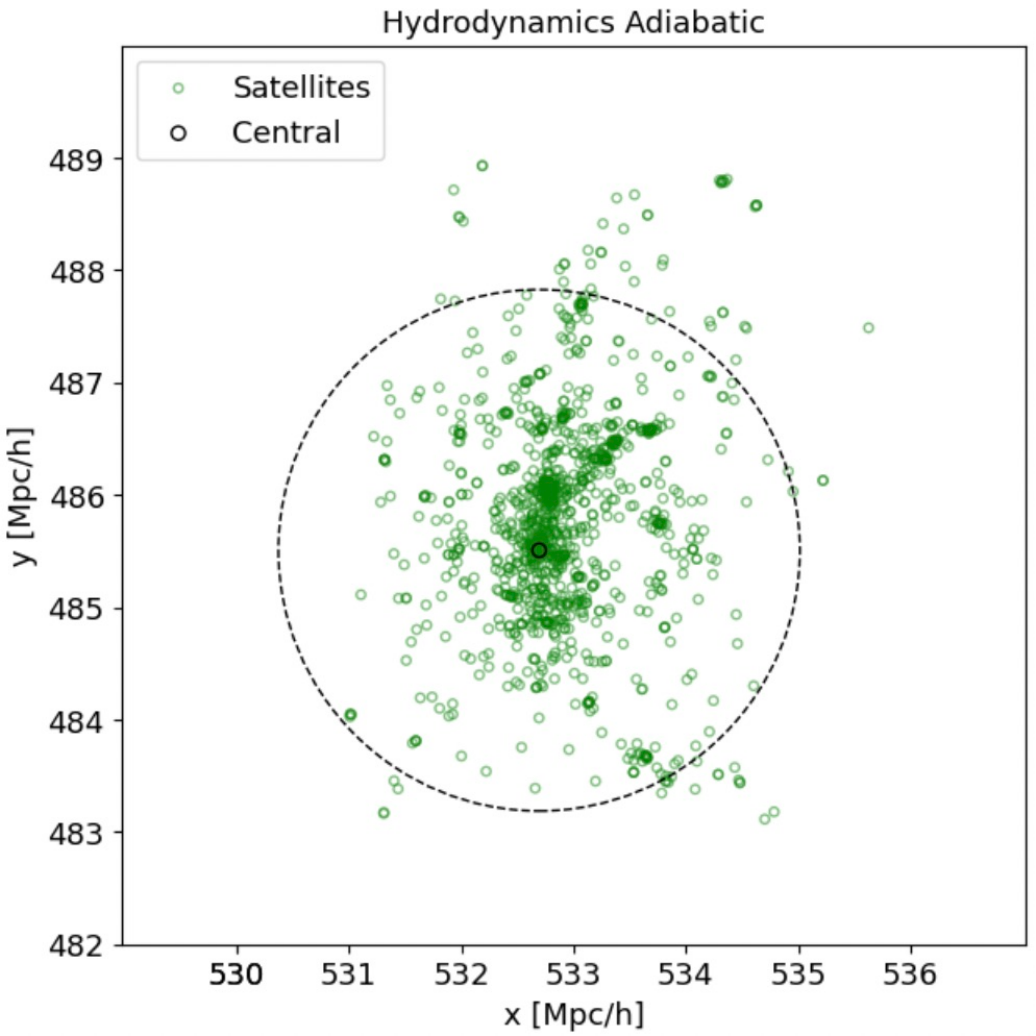}\includegraphics[width=0.29\textwidth]{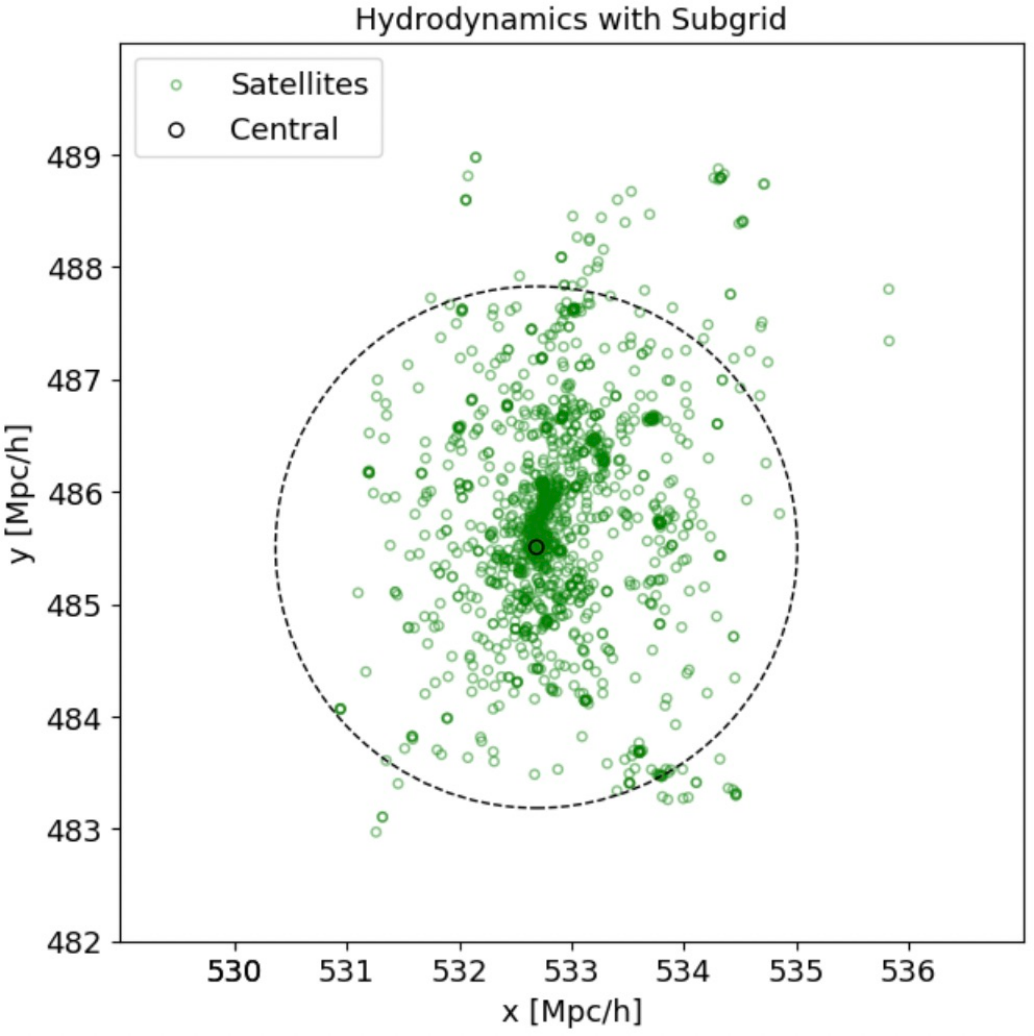}

    \includegraphics[width=0.29\textwidth]{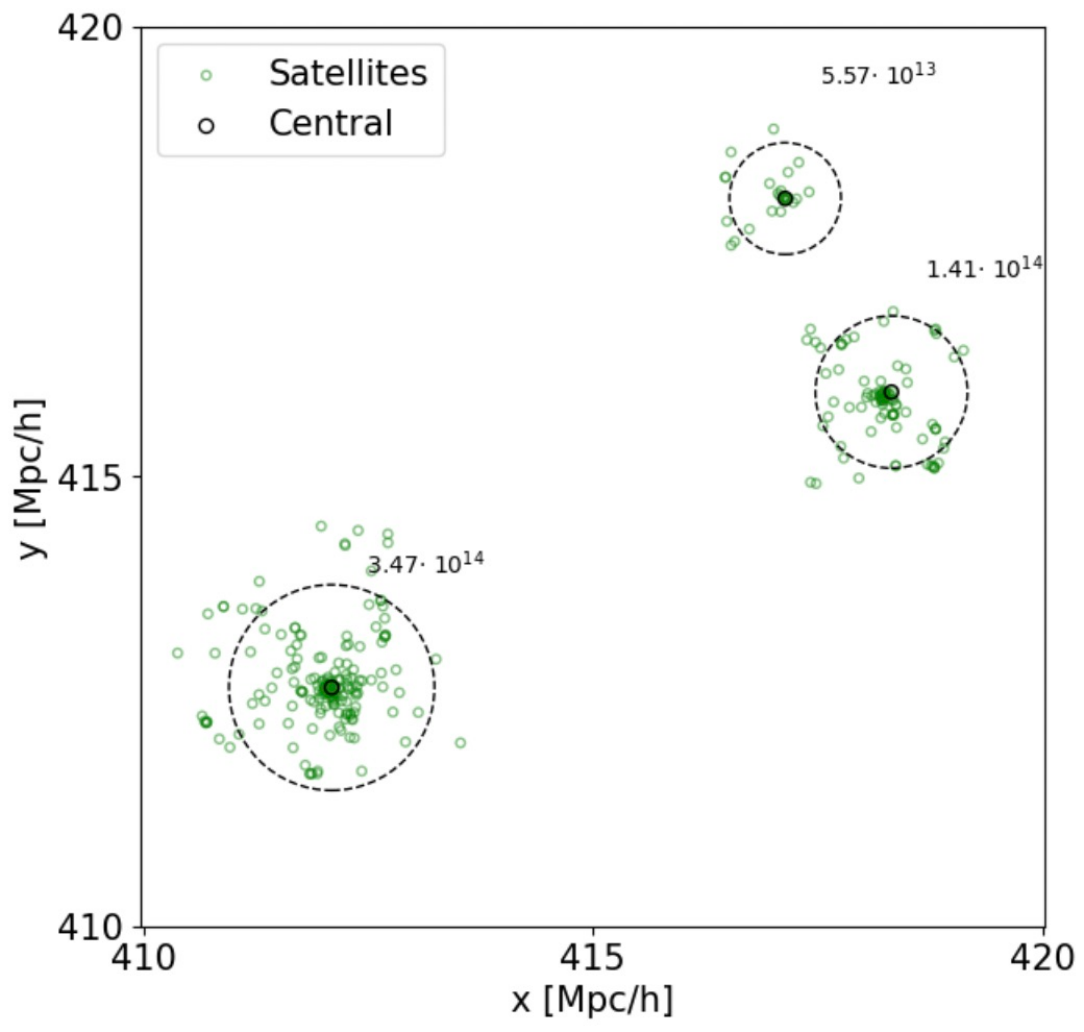}\includegraphics[width=0.29\textwidth]{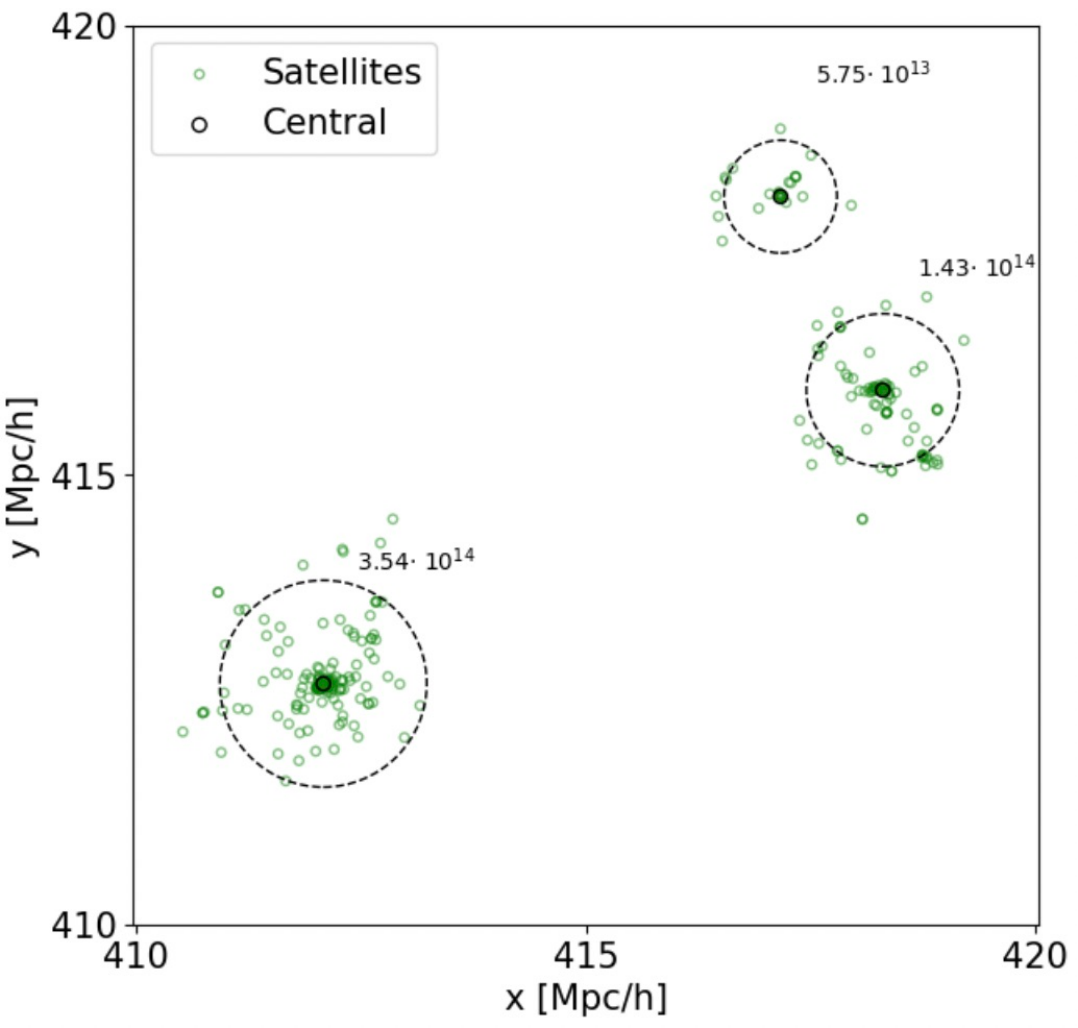}\includegraphics[width=0.29\textwidth]{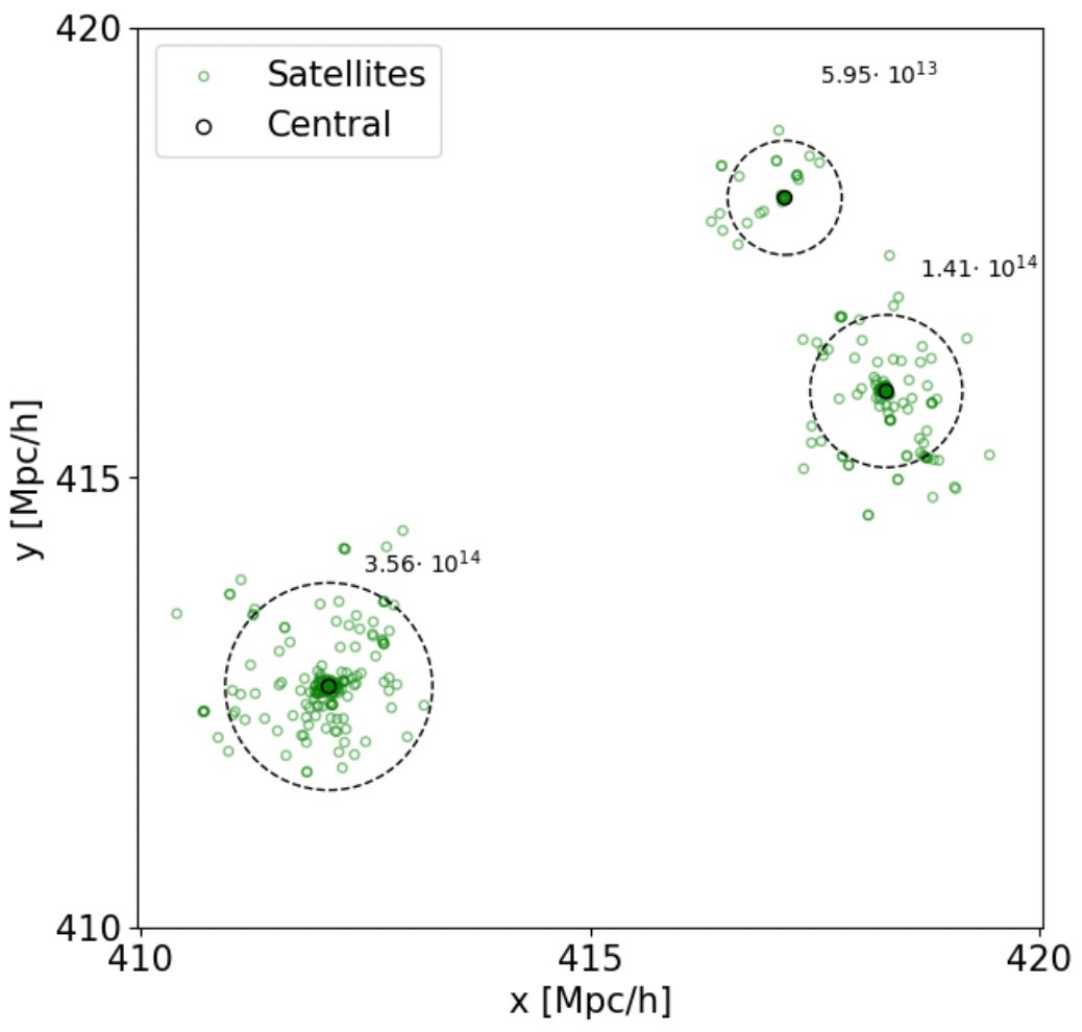}

\vspace{-0.3cm}
  \end{center}
  \caption{Spatial distribution of central and satellite cores in the most massive halo (upper row) and three large halos (lower row) in each simulation. 
  Each halo contains a central core in its most inner region, marked by a black circle. 
  The green circles represent satellite cores. The dashed line indicates the SO halo radius. 
  Note that since we use FoF halos to identify cores, satellite cores may be found outside of the SO radius.
  }
    \label{fig:cores_sims}
\end{figure*}

After finding good statistical agreement of the core populations across the different simulations, we now discuss the spatial distribution of cores. To examine the distribution of cores within halos, we present a simple visual comparison between the different simulations. We show the same largest halo in each simulation in the upper row of Fig.~\ref{fig:cores_sims} and three smaller cluster-sized halos that reside in the same region in the lower row.
The halo masses for the halos shown in the upper row of the figure are 2.90$\times 10^{15}$\,$h^{-1}$M$_\odot$, 2.91$\times 10^{15}$\,$h^{-1}$M$_\odot$, and 2.91$\times 10^{15}$\,$h^{-1}$M$_\odot$, for gravity-only, adiabatic, and hydrodynamics respectively. 
The Euclidean distances between the halo centers of the gravity-only and hydrodynamics simulations are 103\,$h^{-1}$kpc and 97\,$h^{-1}$kpc for the adiabatic and subgrid hydrodynamics simulations, respectively. The distance between the halo centers for the adiabatic and subgrid hydrodynamics simulation is 12\,$h^{-1}$kpc. The dashed line indicates the SO halo radius. There are  4400, 4457, and 4556 satellite cores (green circles) for gravity-only, adiabatic, and subgrid hydrodynamics, respectively.  Note that, since we use FoF halos to identify cores, satellite cores can be found outside of the SO radius. 

Overall, if we take a ``bird's-eye'' view of the core distributions in this figure, it is clear that the substructures are strongly correlated from one panel to the next. Upon closer inspection however, the individual cores are displaced from each other, or even missing, from one simulation to the next. This is a manifestation of the so-called ``butterfly effect'' \citep{genel2019}, where minute perturbations between the simulations grow over time and manifest as macroscopic differences in the positions of objects. In Sec.~\ref{subsec:core_counts}, we noted that the subgrid hydrodynamical core counts are up to 20\% higher than those for the adiabatic and gravity-only simulations and this increased density of cores is visible in the right-hand panel of Fig.~\ref{fig:cores_sims}. Across the simulations, here, as in Fig.~\ref{fig:core_count_beforecut}, the radial distributions of cores are qualitatively similar, but the exact orbital positions are more disparate. This can be understood in terms of the variations in the trajectories of infallen cores due to the additional baryonic effects included in the hydrodynamical simulations. From Fig.~\ref{fig:CoreRadius_dist2} we established that infallen cores have similar histories across the simulations, so it is expected that the structures will align radially, albeit with some minor perturbations due to the aforementioned physics effects.

At large scales, the spatial distributions of cores will be determined by their host halo positions. As the simulations are paired (i.e. have the same initial conditions), we expect that these halos will match between the simulations. As a confirmation of this, we spatially match the halos within their mass bins. 
First, we select the hydrodynamics halos from one of the three highest mass bins in Tab.~\ref{tab:cores_halo}; the gravity-only halos are chosen from corresponding mass bins with slightly relaxed lower and upper limits of $\log_{10}[$M$_h/(h^{-1}$M$_\odot)]$ equal to 11.8, 12.8, and 13.8 for the lower limits and 12.5 and 13.5 for the upper limits (for the 2 lowest mass bins). This widening of the search mass bin accounts for possible fluctuations in the halo masses between the hydrodynamics and the gravity-only simulations. \footnote{We tested the effect of widening the mass bins by choosing different values for the upper and lower limits of the bins and selecting the lowest and highest values, respectively, that resulted in stable matching fractions for the halos across the simulations.} We then match the halos based on their FoF halo center using a k-d~tree algorithm to identify up to 20 closest halo neighbors within a specified distance (ranging from 5\,$h^{-1}$kpc to 300\,$h^{-1}$kpc) of the hydrodynamics halo position. We expect the fraction of matched halos to increase with the specified distance. From the 20 nearest neighbors, we then select the closest match in mass. At 300\,$h^{-1}$kpc, the percentages of subgrid hydrodynamics halos matched with gravity-only ones are 98.2\%, 98.1\%, and 95.8\% for the highest to lowest mass bins, respectively. For the adiabatic hydrodynamics halos matched with gravity-only halos, the corresponding percentage of matches are 98.4\%, 98.3\%, 96.2\%.  In summary,  we can expect therefore, that halos across the simulations will match up spatially (with some displacement) and that structures within those halos will be strongly correlated, but that individual cores within halos may not necessarily match up equally well. 

\section{Cores and Galaxies}
\label{sec:coresandgalaxies}

\begin{figure*}
    \centering
    \includegraphics[width=0.33\linewidth]{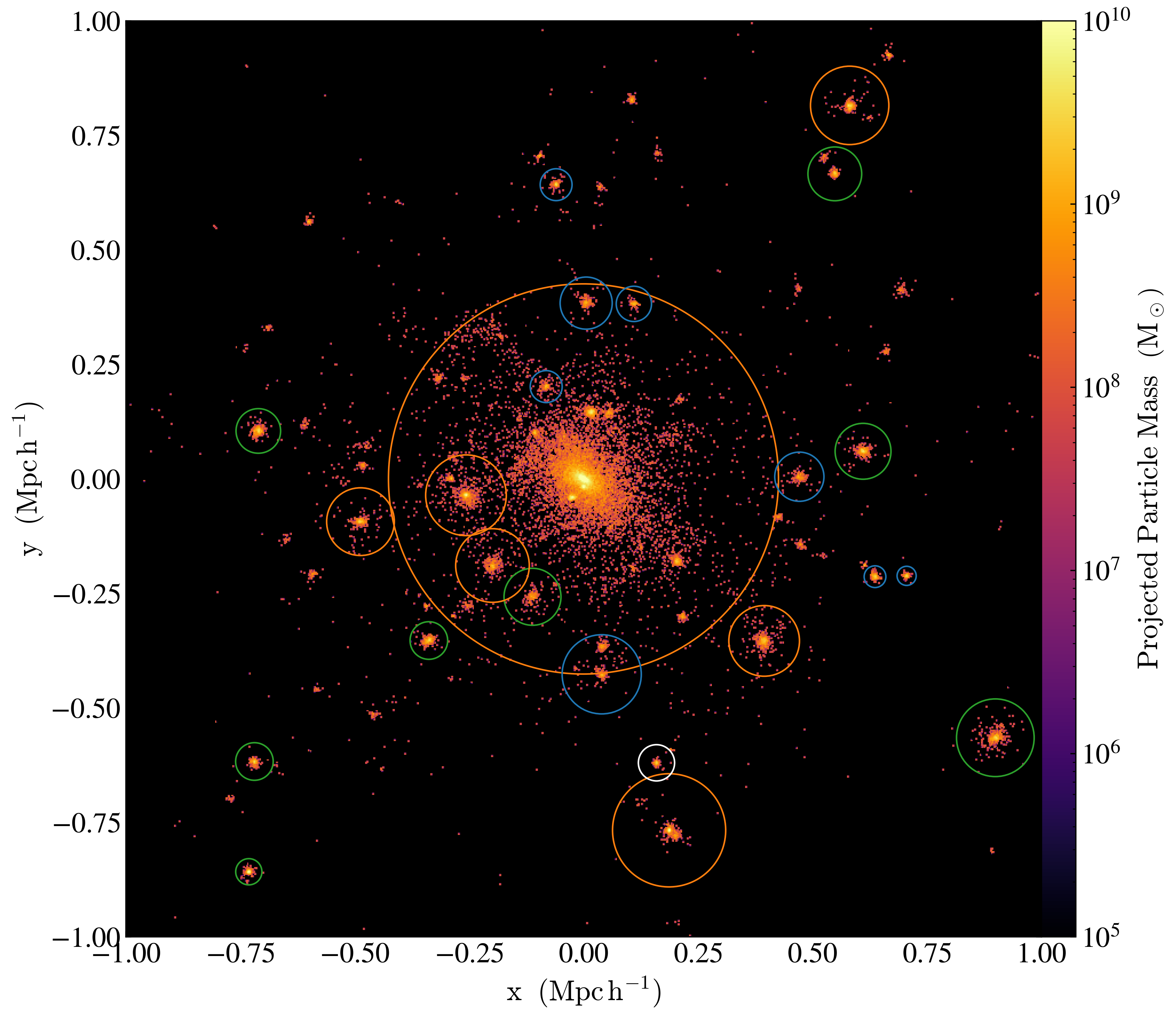}
    \includegraphics[width=0.33\linewidth]{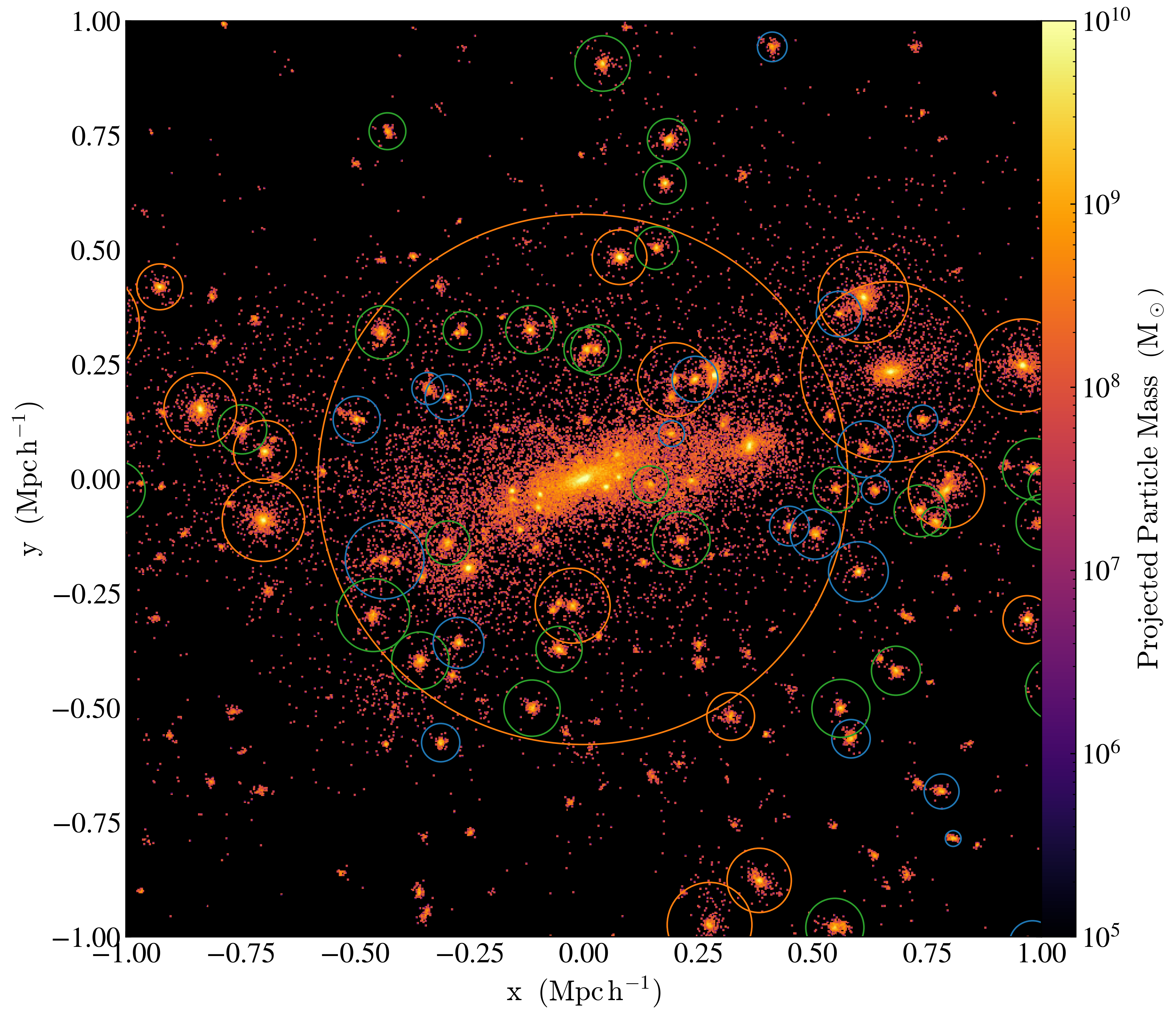}
    \includegraphics[width=0.33\linewidth]{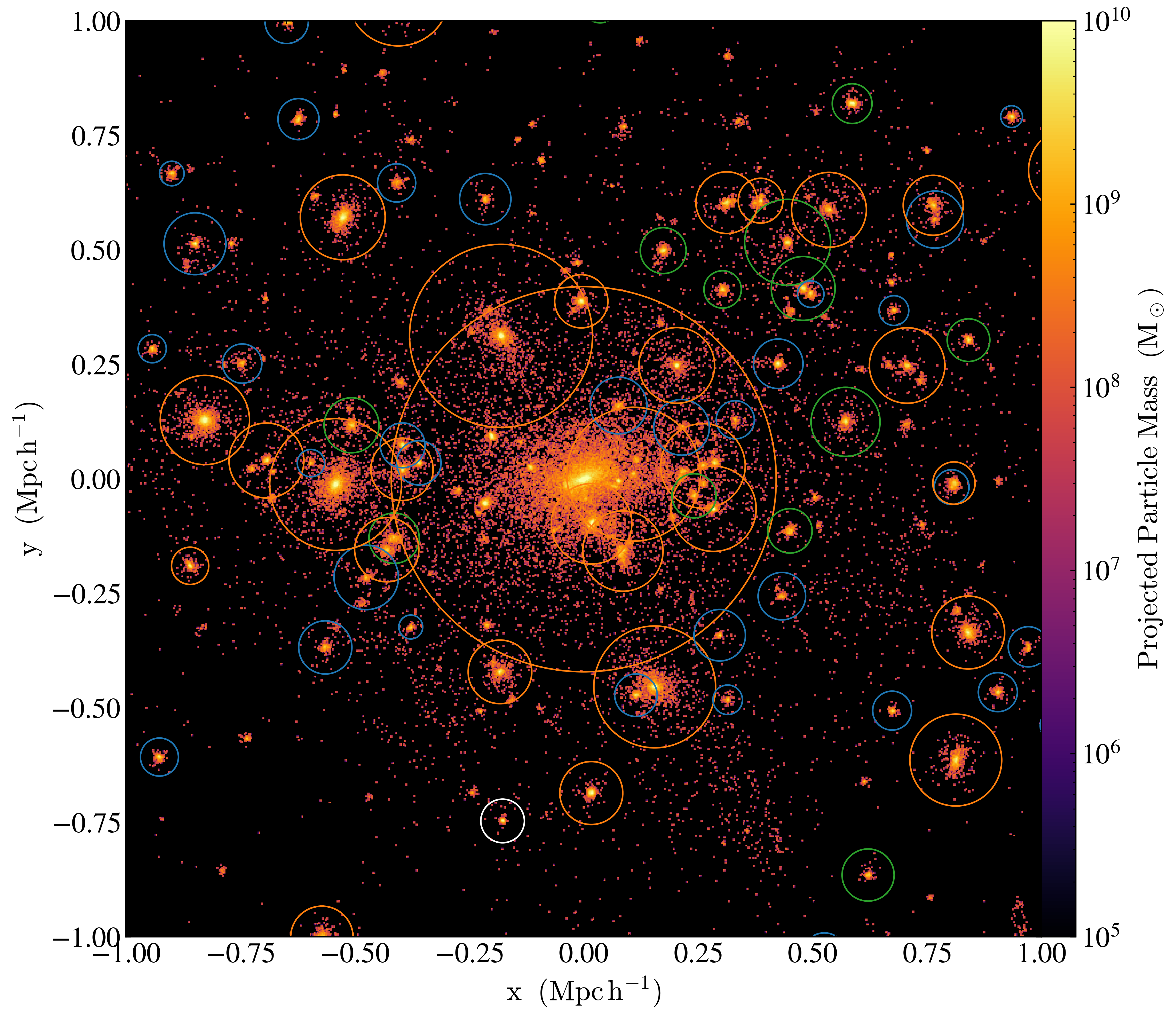}
   \caption{Two-dimensional projection of the mass in a small cut-out region of the subgrid hydrodynamics simulation overlaid with circles whose centers show the location of galaxies identified by the DBSCAN clustering algorithm described in Sec.~\ref{sec:hydro}. The  size of each circle shows the DBSCAN galaxy radius. The colors indicate the number of cores residing within that radius (white $=$ 0, blue $=$ 1, green $=$ 2, orange $\geq$ 3). The same color coding is used in Fig.~\ref{fig:galaxy_vis}.
   \label{fig:vis_gal_core}}
  \end{figure*}

Having established good agreement of the core statistics and radial distributions across the three different simulations, we now turn to the connection between cores and galaxies. As described in Sec.~\ref{sec:hydro}, the hydrodynamics simulation including subgrid models delivers galaxy catalogs, allowing us to study the galaxy-core connection directly. Our goals in this section are threefold: 1) we match cores and galaxies within the hydrodynamics simulation itself to gain insights into the galaxy-core connection directly (Sec.~\ref{sec:core-galaxies}), 2) we develop a methodology based on merging cores into core groups and sub-selecting these groups to achieve a one-to-one correspondence between the core groups and galaxies in the hydrodynamics simulation (Sec.~\ref{sec:pruning}), and 3) we apply the methodology to the gravity-only simulation (Sec.\ref{sec:gravity_only_galaxies}). 

\begin{figure}[t]
  \centering
    \includegraphics[width=3in]{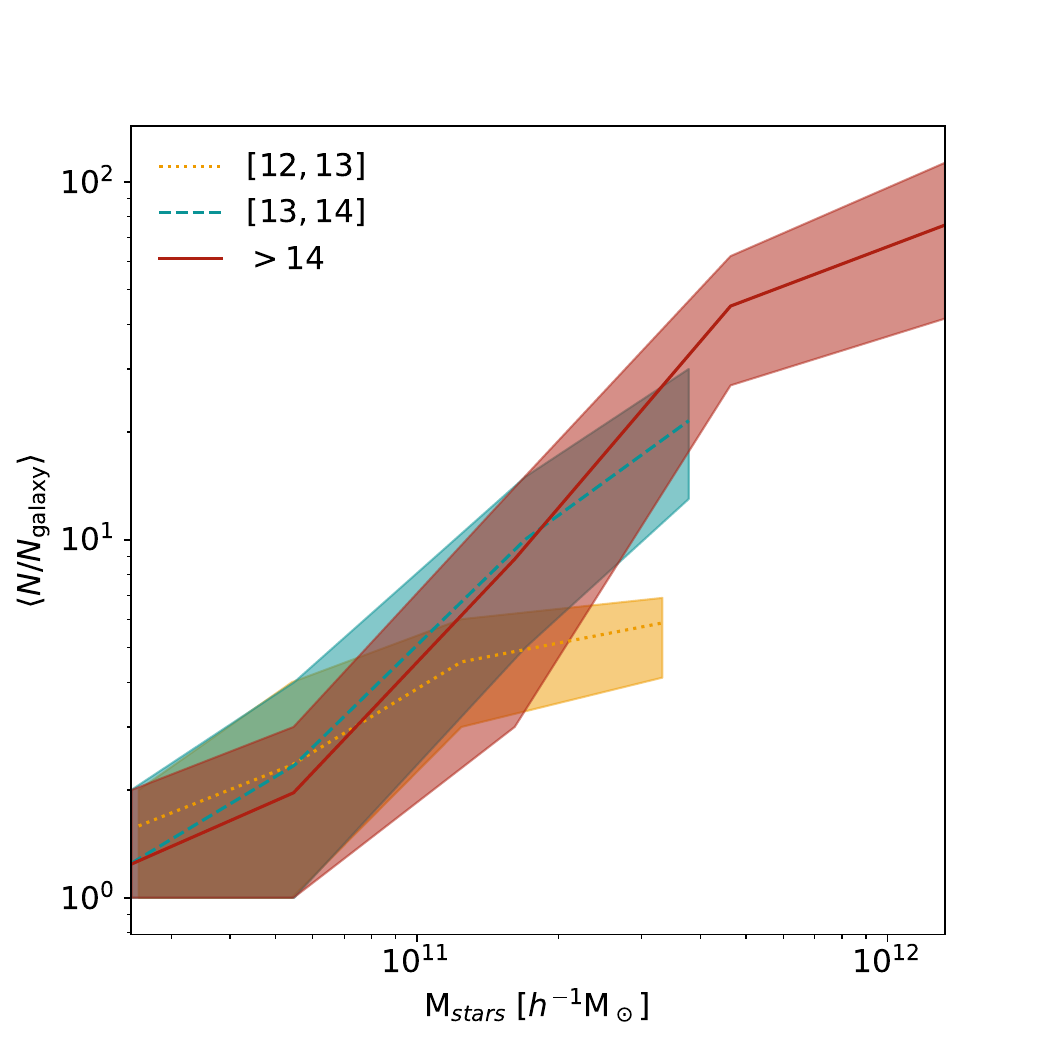}
  \caption{Relationship between the average number of cores within galaxies and their aperture stellar masses, as defined in Sec.~\ref{sec:hydro}. The different colors show the results for galaxies residing in halos in the three highest halo mass bins from  Tab.~\ref{tab:cores_halo}. The lines and shaded bands show the means and the 16th and 84th percentiles, respectively, for the number of cores within each galaxy as a function of the galaxy stellar mass. There is a clear trend showing that the number of cores associated with a galaxy increases with increasing galaxy stellar mass.
  \label{fig:N_vsMstar}}
\end{figure}

\subsection{Matching Cores with Galaxies}
\label{sec:core-galaxies}

In this section, we investigate how well cores and galaxies match within the hydrodynamics simulation itself. We start with a visual inspection. Figure~\ref{fig:vis_gal_core} shows examples for the mapping of cores and galaxies in a small cut-out region of the subgrid hydrodynamics simulation. The galaxies are identified by the DBSCAN algorithm as defined in Sec.~\ref{sec:hydro}. Each circle in the projected mass density distribution marks the location of an individual galaxy, with a radius equal to the galaxy's DBSCAN radius. The color of the circle indicates how many cores are present within that radius: zero cores in white circles, one in blue, two in green and three  or more in orange. The masses of the halos are 3.38, 6.34 and 6.63 $\times 10^{14}$ $h^{-1}M_\odot$ from left to right, respectively. The images confirm that, in general, overdensities in mass, and therefore cores, are excellent indicators for galaxy positions.

Next, we study the core-galaxy connection quantitatively via spatial matching of galaxies and cores in the subgrid hydrodynamics simulation.  We use the information about each galaxy's central position and radius to determine if the galaxy contains a core and to calculate the Euclidean distance between cores and nearby galaxy centers. Within a given halo, we label the galaxy having the largest stellar mass (which in most cases will correspond to the brightest cluster galaxy) as the central galaxy; all others are labeled as satellite galaxies. We note that this definition of the central galaxy is chosen to be compatible with observations; however, it means that the central galaxy may not align with the halo center. The matching proceeds as follows: for each core within the FOF halo, we determine if its position is within a galaxy's DBSCAN radius. If so, we assign that galaxy ID to the core. If a core resides within the radii of multiple galaxies, it is assigned to the closest one. Table~\ref{tab:galaxies_without_cores} shows the percentages of central and satellite galaxies without a core within their DBSCAN radius (coreless galaxies) at different redshifts. Overall, we find that most galaxies contain at least one core, with only 0.28\% of all galaxies (including both centrals and satellites) not associated with a core at $z=0$. The fractions of coreless central and satellite galaxies are only 0.09\% and less than 1\%, respectively.
We have confirmed that these galaxies are well-resolved and gravitationally bound, as expected given that we exclude any objects with fewer than 100 star particles. The rare appearance of coreless galaxies arises from numerical edge cases or thresholding in the core identification and tracking procedures. Since all of these instances represent a tiny fraction of the total population, we consider the small non-zero fraction of coreless galaxies to be reasonable and unlikely to significantly affect our analysis.


\begin{table}[t]
  \begin{center}
  \caption{Central and satellite galaxy counts and percentages of coreless galaxies }\label{tab:galaxies_without_cores}
    \begin{tabularx}{\linewidth}{Xccccccc}
     Redshift  & Coreless & Total     & Coreless   & Total \\
              & Satellites & Satellites & Centrals  & Centrals\\
      \midrule

        0      & 0.82\% & $3.54 \times 10^{5}$  & 0.091\%  & $9.94 \times 10^{5}$ \\
        0.2    & 0.59\% & $3.48 \times 10^{5}$  & 0.094\% & $1.07 \times 10^{6}$ \\ 
        0.5    & 0.49\% & $3.07 \times 10^{5}$  & 0.096\% & $1.08 \times 10^{6}$ \\ 
        1.0    & 0.54\% & $1.95 \times 10^{5}$  & 0.081\% & $9.21\times 10^{5}$ \\ 
        2.0    &0.54\% & $3.99 \times 10^{5}$  & 0.045\% & $3.92\times 10^{5}$ \\
        
    \end{tabularx}
  \end{center}
  {\sc Tab.~\ref{tab:galaxies_without_cores}.--} Fraction of central and satellite galaxies without a core within their DBSCAN radius (coreless galaxies) for different redshifts in the subgrid hydrodynamics simulation. We find that most galaxies without cores are satellite galaxies.
\end{table}

We further investigate the relationship between a galaxy's stellar mass and the number of cores contained within the galaxy. Figure~\ref{fig:N_vsMstar} shows the number of cores falling within the galaxy's radius as a function of its stellar mass for the three highest SO halo mass bins from Tab.~\ref{tab:cores_halo}. The lines and shaded bands show the means and the 16th and 84th percentiles of the core counts per galaxy, respectively. There is a strong correlation between the core counts and the stellar mass with only modest dependence on the mass of the host halo. Hence, clusters of cores can be used not only to locate galaxies but also to provide information, albeit with large scatter, about galaxy stellar masses. We defer the details of calibrating such estimates to future work. 

\subsection{Core Pruning}
\label{sec:pruning}
In this section, we match groups of cores with galaxies in the subgrid hydrodynamics simulation to derive a merging criterion for the cores. We have already established that we have more cores than galaxies, and that more than one core may reside within a galaxy radius. Therefore, we develop a methodology to merge these cores, so that, on average, there is one core group per galaxy. This methodology can then be applied to the gravity-only simulation to reproduce statistically the expected distribution of galaxies obtained from a subgrid hydrodynamics simulation with similar resolution.

To group the cores, we use a standard FOF finder to cluster neighboring cores with linking-length values of 0.06, 0.08, 0.1, and 0.12\,$h^{-1}$Mpc.  Core groups assembled using larger linking lengths will contain more cores and be fewer in number, compared with those resulting from using smaller linking lengths. Once the cores have been identified as part of a core group, we merge them and choose the core with the largest infall SO halo mass to represent the mass of the group, thus pruning the core data. This procedure groups only neighboring cores together and does not make use of individual particle positions within cores.


\begin{figure}[t]
  \centering
    \includegraphics[width=3in]{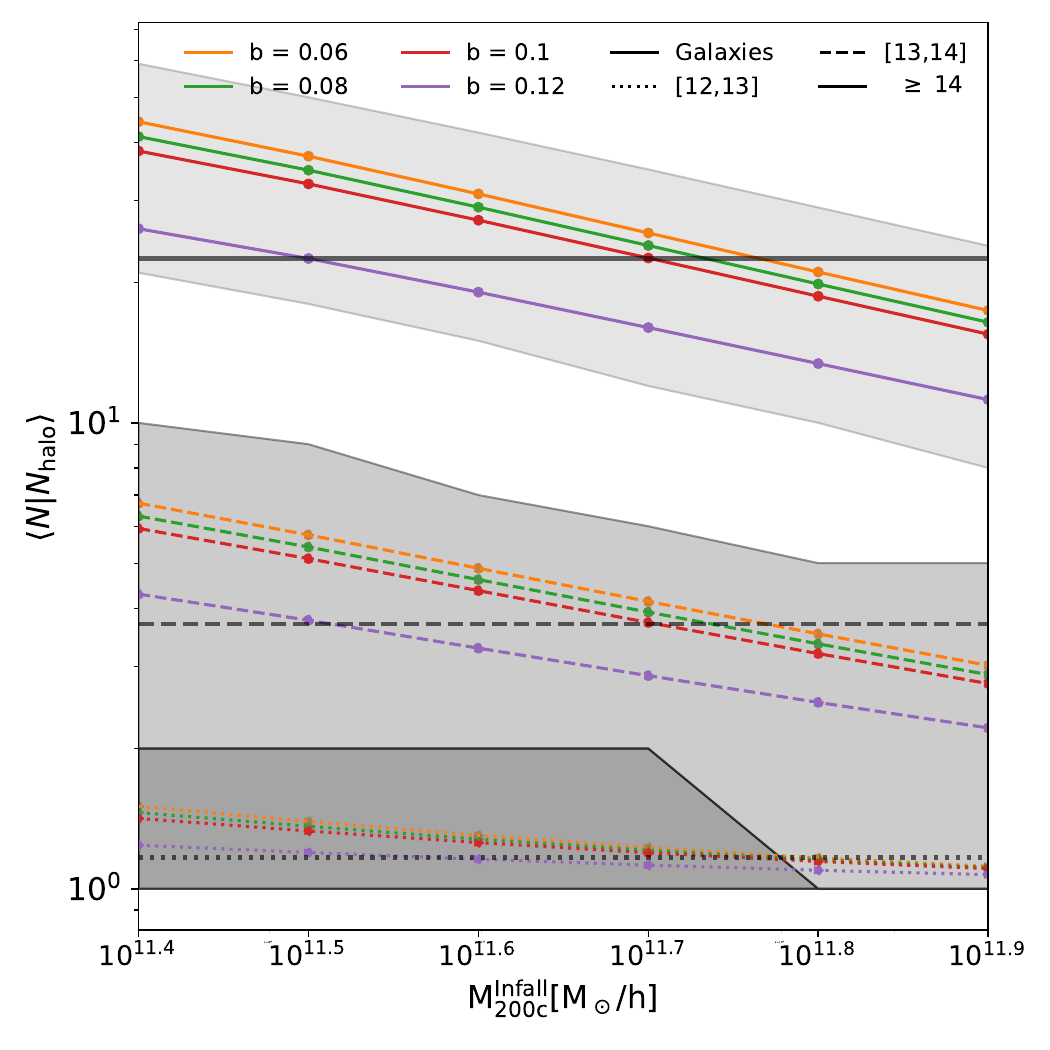}
  \caption{Average number of core groups per halo for different merging criteria and infall dark matter mass cuts. The colored lines show the results for different linking lengths $b$ and the shaded bands show the 16th and 84th percentiles for  $b=0.08$\,$h^{-1}$Mpc.  The black lines show the average number of galaxies per halo for each halo mass bin.}
  \label{fig:N_vs_infallcut_100}
\end{figure}

To match cores to galaxies, in addition to grouping the cores as described above, we also need to apply an infall halo mass cut to the cores. Measurements of the Stellar-Mass to Halo-Mass (SMHM) relation \citep{behroozi2019universemachine}, show that our cut on galaxy stellar masses of $1.87 \times 10^{}$\,$h^{-1}$ M$_\odot$ correspond to values of $\log_{10}(\mathrm{M}_{h}/$M$_\odot) \approx 11.7 - 11.8$. Hence, we expect that we need to consider only cores in halos whose infall masses are higher than this value. Furthermore, at the end of this section,  we compare core and galaxy density profiles. \citet{korytov2023modeling} showed that selecting cores using an infall mass threshold reduces the normalization of the core density profiles significantly while keeping the shape intact. Therefore, we expect that applying an infall mass cut on the cores should simply lower the core-density profiles but not affect which choice of linking length provides the best agreement.

\interfootnotelinepenalty=10000

To refine the choice of the infall mass threshold, we evaluate the average number of core groups per halo, after applying a cut on the infall dark matter mass for each group.\footnote{The dark matter mass in this case is an FoF mass that excludes contributions from halo fragments that were overlinked by the halo finder (see~\citealt{rangel2017building}).} Figure~\ref{fig:N_vs_infallcut_100} shows the average number of core groups, merged with various linking lengths (colored lines) and selected according to the infall mass threshold given on the horizontal axis. The gray shaded bands show the 16th and 84th percentiles of the number of core groups for a linking length value of $b=0.08$\,$h^{-1}$Mpc. The black lines show the average number of galaxies per halo, which are  1.61, 7.33, and 46.8 for the lowest to highest mass bins, respectively.
The number of core groups decreases with increasing infall mass threshold for all mass bins and all linking lengths. The points at which the colored lines cross the black lines give the optimum value for the infall mass threshold for each mass bin.  
These points lie at about $10^{11.73}$\,$h^{-1}$M$_\odot$ for each mass bin for  $b=0.08$\,$h^{-1}$Mpc, and fall slightly above and below this range for lower and higher linking lengths, respectively.

As we discuss in the following, we choose an infall mass cut of $10^{11.73} = 5.4\times 10^{11}$\,$h^{-1}$M$_\odot$, leading to good qualitative agreement between the numbers of core groups and galaxies and the shapes and normalizations of their density profiles for our preferred choice of $b=0.08$\,$h^{-1}$Mpc.
In Tab.~\ref{tab:galaxies_and_core_groups}, we provide a summary of the total number of core groups found for different linking lengths and the three highest SO halo mass bins from Tab.~\ref{tab:cores_halo}, before and after the infall mass cut.  We also provide the total number of galaxies for these mass bins. The best agreement between these core and galaxy statistics is found for $b=0.08$\,$h^{-1}$Mpc and the infall mass cut of $\log_{10}(\mathrm{M}_{h}/$M$_\odot h^{-1}) = 11.73$ which we apply to the remainder of the results presented in this section.

\begin{table}[ht]
  \begin{center}
  \caption{Comparison of the number of galaxies and core groups} \label{tab:galaxies_and_core_groups}
    \begin{tabularx}{\columnwidth}{Yccc}
    \raggedright $\log_{10}$(M$_h/(h^{-1}$M$_\odot))$ & $[12, 13]$ & $[13, 14]$ & $\geq 14$ \\
     \cmidrule(r){2-4}\\
     & \multicolumn{3}{c}{Galaxies} \\
    \cmidrule(r){2-4} 
     & $6.39 \times 10^{5}$ & $2.29 \times 10^{5}$  & $8.13 \times 10^{4}$ \\[0.3cm]
    $b$ [$h^{-1}$Mpc] & \multicolumn{3}{c}{Core Groups without Infall Mass Cut} \\
     \cmidrule(r){2-4} 
    0.06 &$9.35 \times 10^{5}$ & $5.36 \times 10^{5}$ & $2.15 \times 10^{5}$\\
    0.08 & $8.96 \times 10^{5}$ & $4.95 \times 10^{5}$ & $1.96 \times 10^{5}$\\
    0.10 & $8.60 \times 10^{5}$& $4.61 \times 10^{5}$ & $1.80 \times 10^{5}$\\
    0.12 & $7.19 \times 10^{5}$& $ 3.18 \times 10^{5}$ & $ 1.18 \times 10^{5}$\\[0.3cm]
     &  \multicolumn{3}{c}{Core Groups with Infall Mass Cut} \\
     \cmidrule(r){2-4} 
     0.06 &$6.60  \times 10^{5}$ & $2.43 \times 10^{5}$ & $8.73 \times 10^{4}$\\ 
     0.08 & $6.53 \times 10^{5}$ & $2.31 \times 10^{5}$ & $8.22 \times 10^{4}$\\
     0.10 & $6.39 \times 10^{5}$& $2.29  \times 10^{5}$ & $7.74 \times 10^{4}$\\ 
     0.12 & $6.11 \times 10^{5}$& $1.71 \times 10^{5}$ & $5.57 \times 10^{4}$\\
    \end{tabularx}
  \end{center}
  {\sc Tab.~\ref{tab:galaxies_and_core_groups}.--} Number of galaxies and core groups obtained after applying different choices for the linking length $b$. Numbers pertain to the subgrid hydrodynamics simulation for the halo mass bins given in the table.

\end{table}

\begin{figure}[t]
  \centering
    \includegraphics[width=\linewidth]{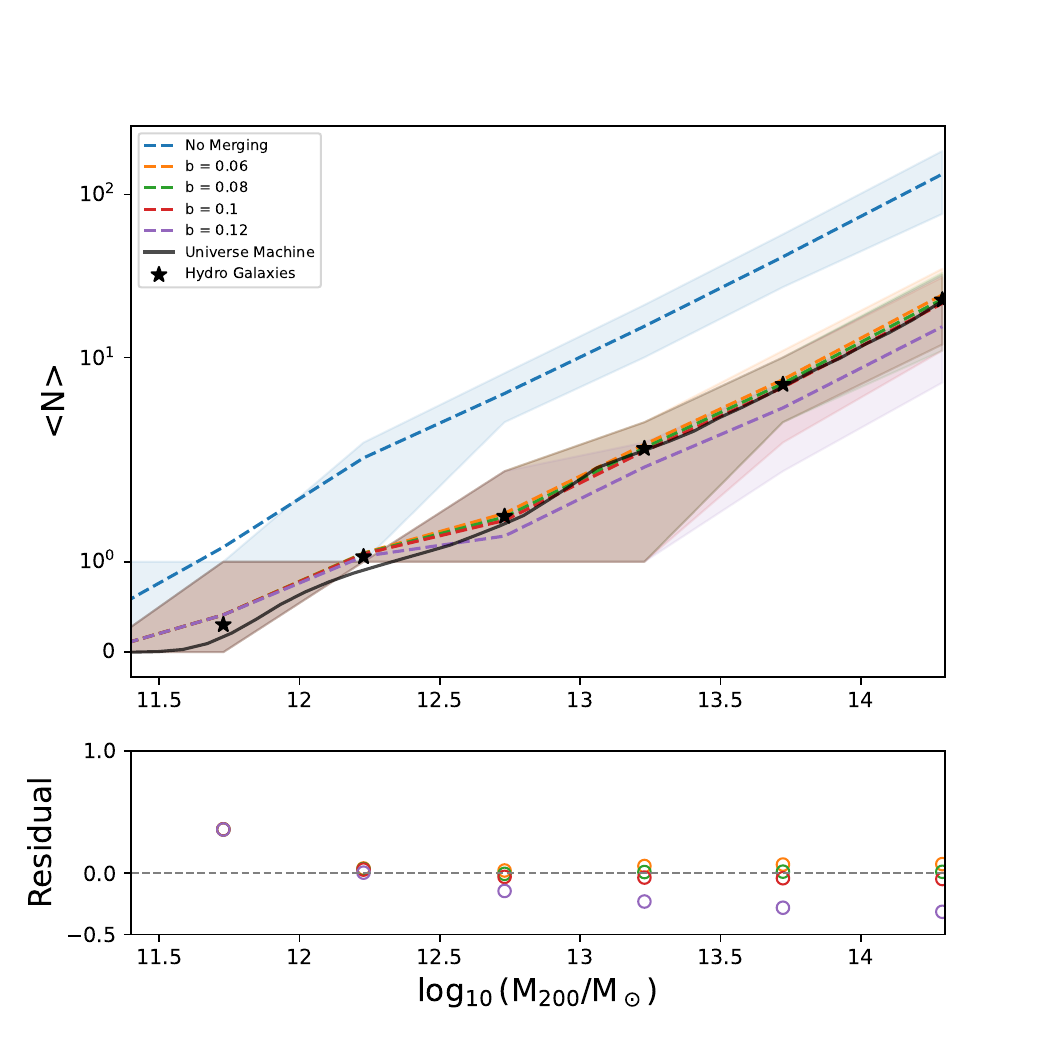}
  \caption{Upper panel: Relationship between the average number of core groups within halos and their host SO halo mass for different merging criteria in the subgrid hydrodynamics simulation. We apply an infall mass cut of $\log_{10}(M_{h}/$M$_\odot h^{-1}) = 11.73$ to the core groups we consider here. The colored dashed lines and shaded bands show the means and the 16th and 84th percentiles, respectively, for the number of core groups. The colors correspond to different linking lengths as given in the legend. The solid black line shows the average galaxy counts per halo from the Universe Machine simulation~\citep{behroozi2019universemachine} after applying a stellar mass cut equivalent to the 100 star-particle cut that we applied to the galaxies in the hydrodynamics simulations. The black stars show the average galaxy counts for the subgrid hydrodynamics simulation. Lower panel: Residuals between the average numbers of cores and galaxies (normalized by the number of galaxies). }
  \label{fig:N_vs_M200_allCores}
\end{figure}

Figure~\ref{fig:N_vs_M200_allCores} shows the relationship between the average number of core groups per halo and their SO host halo masses for different linking lengths. The dashed colored lines show the results for unmerged cores (blue) and four choices of linking lengths, $b=0.06$\,$h^{-1}$Mpc (orange),  $b=0.08$\,$h^{-1}$Mpc (green),  $b=0.10$\,$h^{-1}$Mpc (red),  and  $b=0.12$\,$h^{-1}$Mpc (purple). The solid black line and black stars show the average number of galaxies per halo from Universe Machine~(UM)~\citep{behroozi2019universemachine} after applying a stellar mass cut of $1.87\times 10^{10}$\,$h^{-1}$M$_\odot$ to the galaxies in the UM and the subgrid hydrodynamics simulations, respectively. There is good agreement between the galaxy counts and core counts for the three smallest choices of linking lengths. Careful examination of the residuals shows that for the smallest linking-length of $b=0.06$\,$h^{-1}$Mpc, the average number of core groups slightly exceeds the average number of galaxies from the subgrid hydrodynamics simulation, indicating an undermerging of cores. As we increase the value of $b$, we find excellent agreement of core group counts with galaxy counts at $b=0.08$\,$h^{-1}$Mpc. For higher values of $b \geq 0.1$\,$h^{-1}$Mpc, there is a deficit of core group counts compared to galaxy counts, indicating an overmerging of cores. We studied a wide range of linking lengths and found that grouping cores with linking lengths of $b < 0.08$\,$h^{-1}$Mpc or $b > 0.08$\,$h^{-1}$Mpc would, in effect, be overcounting or undercounting galaxies, respectively. From this analysis, we conclude that a linking length of $b = 0.08$\,$h^{-1}$Mpc provides the best match, over a range of halo masses, for the average number of galaxies per halo in the subgrid hydrodynamics simulation. 

\begin{figure*}[t]
  \begin{center}
    \includegraphics[width=0.333\textwidth]{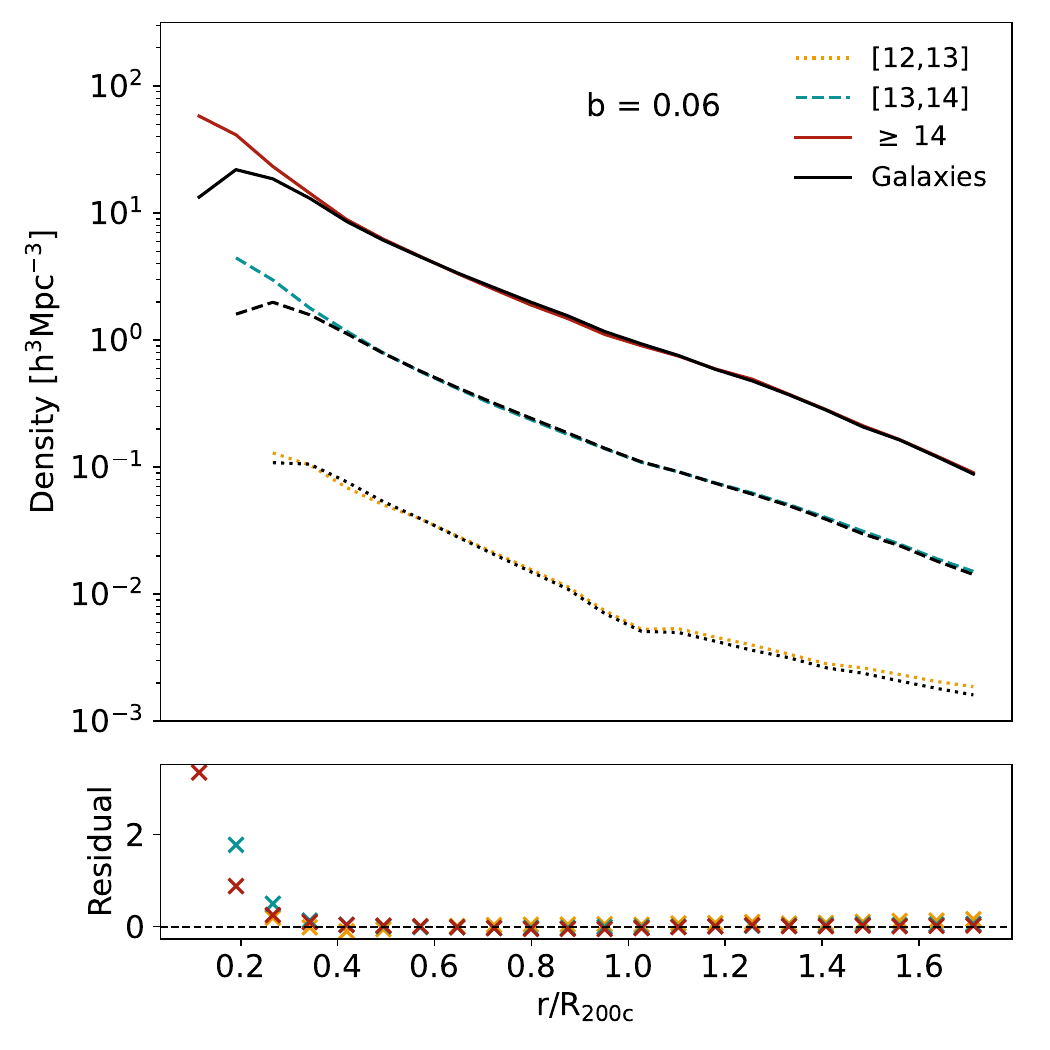}\includegraphics[width=0.33\textwidth]{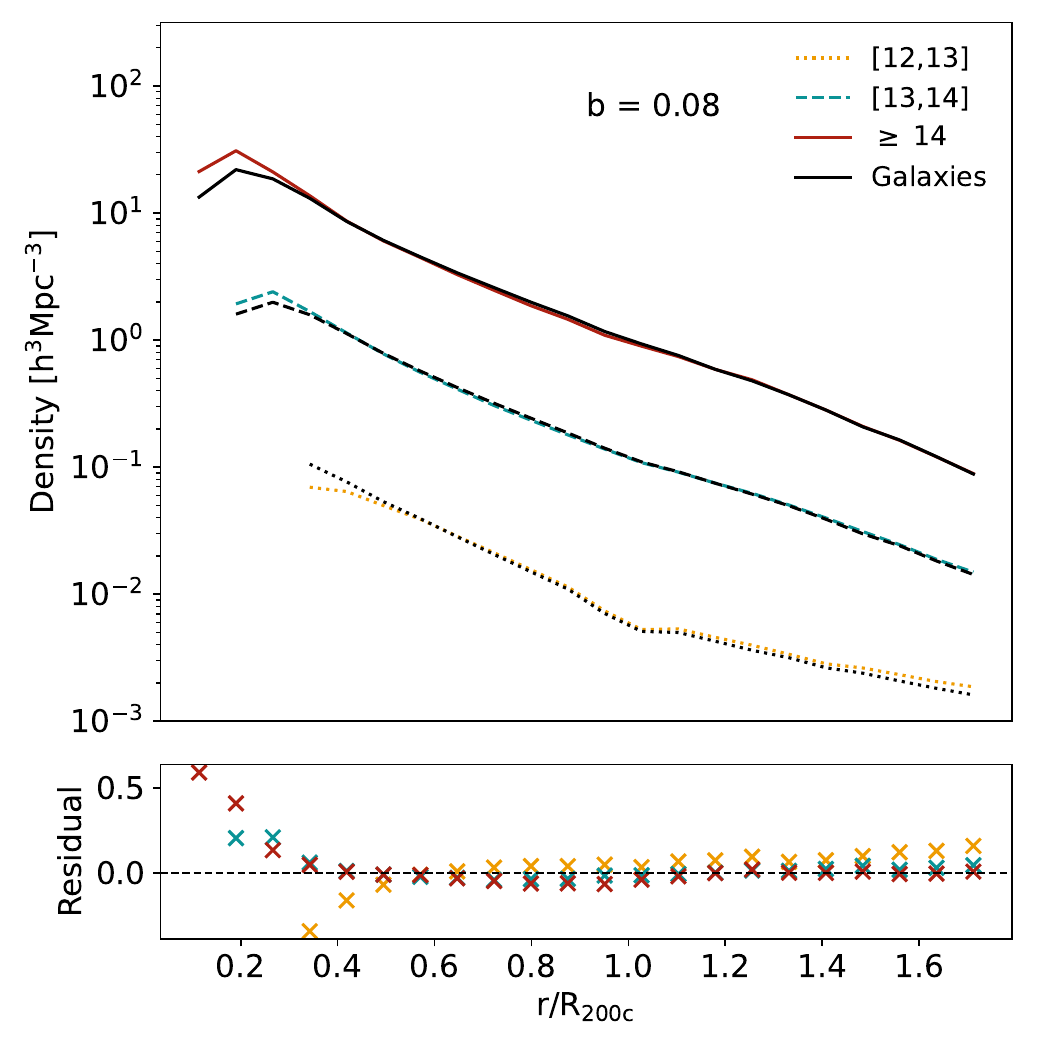}\includegraphics[width=0.33\textwidth]{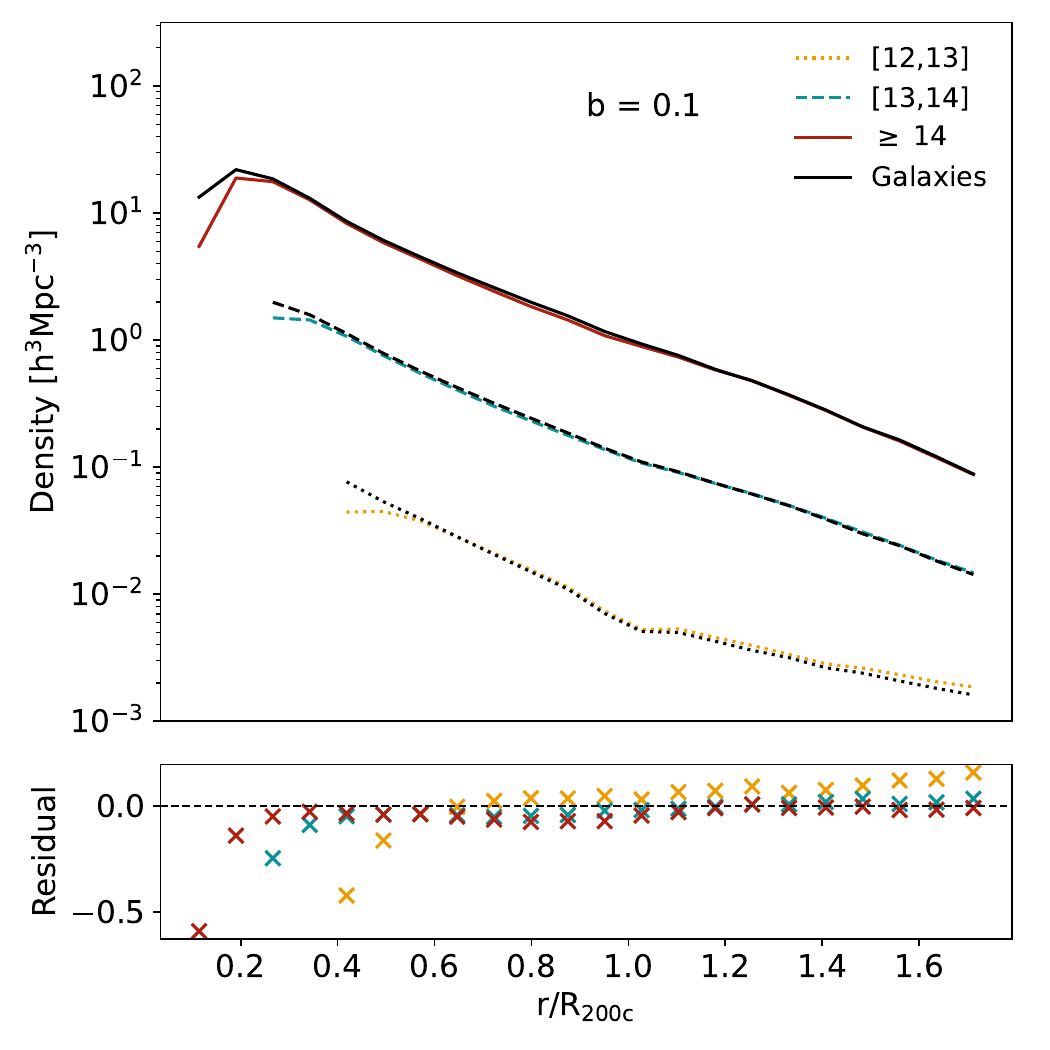}

  \end{center}
  \caption{Comparison of the density profiles for core groups and galaxies (upper panels) and residuals between these profiles (lower panels) in the subgrid hydrodynamics simulation for the three highest SO halo mass bins from Tab.~\ref{tab:cores_halo} (red, cyan, yellow). The left, middle and right columns show the profiles for linking lengths of $b = 0.06$\,$h^{-1}$Mpc, $b = 0.08$\,$h^{-1}$Mpc, and $b = 0.1$\,$h^{-1}$Mpc, respectively. For comparison, the black lines in each plot (solid, dashed and dotted) show the corresponding galaxy density profiles from the subgrid hydrodynamics simulation for each mass bin. We have applied an infall mass cut of $\log_{10}(M_{h}/$M$_\odot h^{-1}) = 11.73$ to the core groups we consider here and have removed points below values of $b/\langle R_{200c}\rangle$, where $\langle R_{200c}\rangle$ is the mean value of $R_{200c}$ for each mass bin.}
 
    \label{fig:Profiles_with_gals_withcut}
\end{figure*}

We can further refine the optimal choice of $b$ for matching core groups with galaxies by investigating the effects of different choices of linking lengths on the density profiles of core groups compared with those of galaxies, going beyond simply matching the counts of core groups and galaxies. In the upper (lower) panels of Fig.~\ref{fig:Profiles_with_gals_withcut}, we show the core group density and galaxy density profiles (residuals between the profiles) as a function of $r/R_{200c}$ for the three highest SO halo mass bins from Tab.~\ref{tab:cores_halo}. The left, middle and right columns of the figure show the results for linking lengths of $b = 0.06$\,$h^{-1}$Mpc, $b = 0.08$\,$h^{-1}$Mpc, and $b = 0.1$\,$h^{-1}$Mpc, respectively. For comparison, the black lines in each plot (solid, dashed and dotted) show the corresponding galaxy density profiles from the subgrid hydrodynamics simulation for each mass bin.\footnote{The slight dip at $r/R_{200c}$ = 1 is an artifact caused by normalizing each core‑to‑halo separation by the host’s own $R_{200c}$. A core located near the halo boundary adds mass, which marginally increases the radius that satisfies the density threshold of $200\rho_{crit}$; when that happens, the same core is recorded at $r/R_{200c} < 1$. While this shift is negligible for any single halo, stacking many halos, especially low‑mass ones, results in an apparent deficit right at $r/R_{200c}$ = 1. The dip therefore reflects the definition of the radial coordinate and our choice of stacking the profiles in  $r/R_{200c}$ units, and not a genuine lack of cores at the (not well-defined) halo edge.} Note that, since the approximate resolution limit for each profile is given by the value of the linking length $b$, we show only values of $r/R_{200c}$ that exceed
$b/\langle R_{200c}\rangle$, where $\langle R_{200c}\rangle$ is the mean value of $R_{200c}$ for each mass bin.

Overall, the shapes of the core group density profiles agree to within 10-20\% with those of the galaxy density profiles except in the innermost regions of the halos for $r/R_{200c} \lesssim 0.3$. Even in these inner regions, both core and galaxy profiles show similar dips due to the effects of the clustering algorithms defining these objects. The differences here are more strongly dependent on the value of the mass bin. 
The core profiles tend to exceed the galaxy profiles in the outer regions of the halo $r/R_{200c} \gtrsim 1.0$, with the most pronounced effect occurring for the lower mass bins.

The residuals in the lower panels provide a more detailed view of the effects of changing $b$. 
For a value of $b=0.06$\,$h^{-1}$Mpc, shown in the left lower panel of the figure, the residuals for the inner regions of the halo for the two highest mass bins are positive and rise sharply for $r/R_{200c} \lesssim 0.25$. We conclude that this linking length leads to an undermerging of cores with respect to galaxies. On the other hand, for $b=0.1$\,$h^{-1}$Mpc in the right lower panel of the figure, the residuals for all the mass bins are negative for $0.05 \lesssim r/R_{200c} \lesssim 0.5$,  indicating that these cores have been overmerged with respect to the galaxy densities. In the middle panel, for $b=0.08$\,$h^{-1}$Mpc, we see both positive and negative differences in the residuals, indicating both undermerging and overmerging, depending on the mass bin. While it is not possible to find a single value of $b$ that leads to perfect agreement between the shapes of the core group and galaxy density profiles for all mass bins, overall the best agreement between the two is achieved for a linking length of $b = 0.08$\,$h^{-1}$Mpc, consistent with the conclusion drawn from Fig.~\ref{fig:N_vs_M200_allCores}.

\begin{figure}[t]
  \centering
    \includegraphics[width=3in]{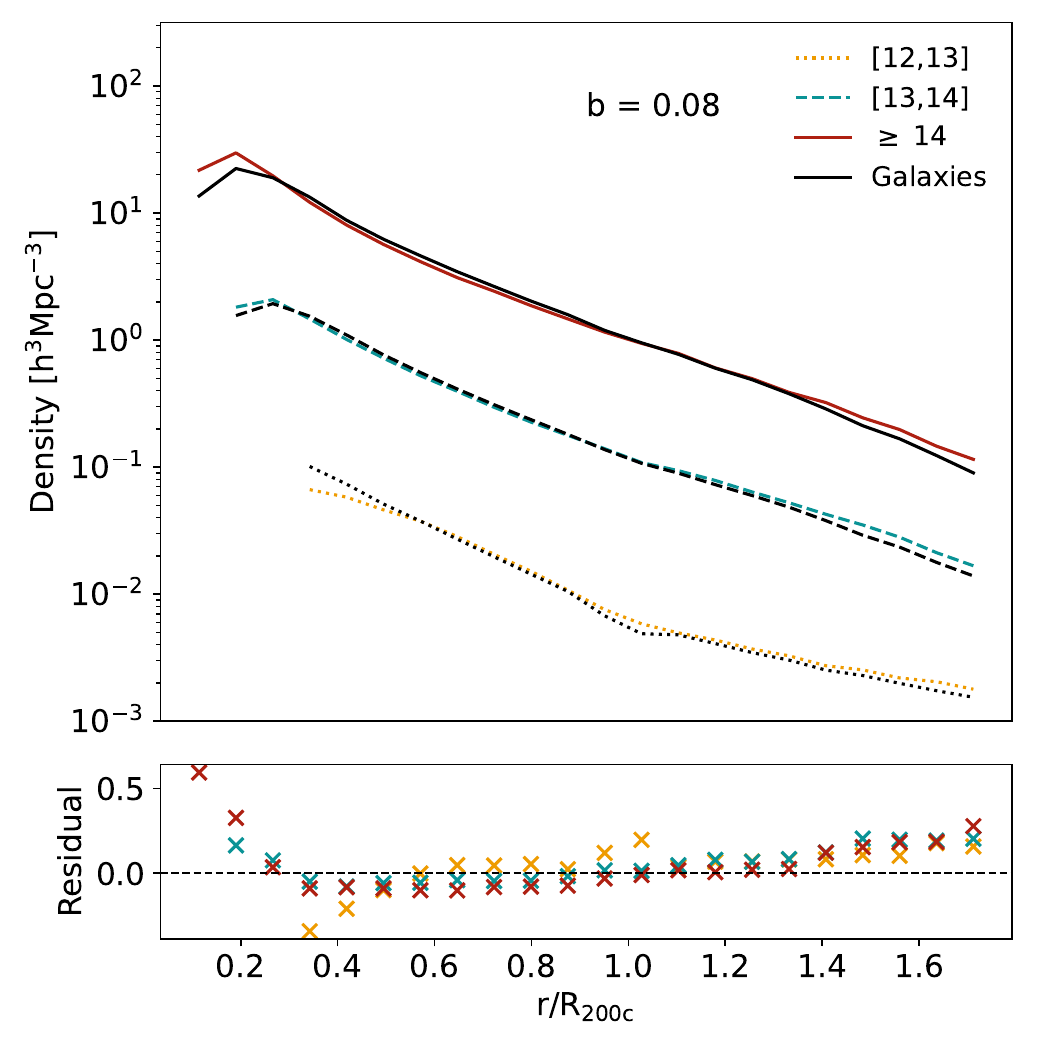}
  \caption{As for Fig.~\ref{fig:Profiles_with_gals_withcut} but with cores from the gravity-only simulation. As with the subgrid cores, the gravity-only cores are selected with a dark matter infall mass cut of $6.38 \times 10^{11}$\,$h^{-1}M_\odot$ (see text) and grouped with a linking length of $b = 0.08$\,$h^{-1}$Mpc. The agreement between the core group and galaxy profiles is only slightly worse than that achieved for the subgrid simulation. In the intermediate regions of the halo  ($0.5 \lesssim r/R_{200c} \lesssim 1.4$) the agreement is within 10-15\% and rises to 20\% in the outer regions ($r/R_{200c} \gtrsim 1.4$).}

  \label{fig:profile_gravity-only}
\end{figure}

\subsection{Gravity-only Cores as Galaxy Proxies}
\label{sec:gravity_only_galaxies}
So far, we have compared the cores from the subgrid hydrodynamics simulation with their associated galaxies to develop a methodology for using cores as galaxy proxies. Since we established in Sec.~\ref{subsec:CoreMatching} that the radial profiles of cores match well across simulations, we can now apply this methodology to the gravity-only simulation. In Fig.~\ref{fig:profile_gravity-only}, we compare the core group density profiles derived from the gravity-only simulation after applying the core-merging criterion and a dark-matter infall mass cut\footnote{The  infall mass cut used in the hydrodynamical simulation has been scaled by the ratio of the dark-matter particle masses in the two simulations.  This scaling accounts for the fact that baryonic masses are not included in the determination of the hydrodynamics infall mass cut.} of $10^{11.80} = 6.38\times 10^{11}$\,$h^{-1}$M$_\odot$ to the galaxy density profiles from the subgrid hydrodynamics simulation. As in the upper panel of Fig.~\ref{fig:Profiles_with_gals_withcut}, the colored lines show the core group densities for three different mass bins and the black lines show the galaxy profiles. The lower panel shows the residuals. The total number of core groups from the lowest to the highest mass bin for a linking length $b=0.08$\,$h^{-1}$Mpc are: $6.77 \times 10^{5}$, $2.35 \times 10^{5}$, $8.12 \times 10^{4}$, respectively.
Comparing these counts to the number of galaxies in Tab.~\ref{tab:galaxies_and_core_groups}, we find that they agree within 6\%  across all mass bins.

The agreement between the core group and galaxy profiles is comparable to that achieved for the subgrid simulation in the intermediate regions of the halo, and is within 10-15\% for $0.5 \lesssim r/R_{200c} \lesssim 1.4$ and rises to 20\% for $r/R_{200c} \gtrsim 1.4$. The agreement for the lowest mass bin is slightly worse than that for the two highest mass bins.

Although the agreement between the core group and galaxy profiles is not quite as good as that shown in Fig.~\ref{fig:Profiles_with_gals_withcut}, the differences are still very modest. Thus, our methodology can be used to populate a gravity-only simulation with a sample of galaxies whose counts and radial density profiles are in broad agreement with those of galaxies from a hydrodynamics simulation.

\section{Summary and Outlook}
\label{sec:conclusion}

In this paper, we have used three 576\,$h^{-1}$Mpc companion simulations -- gravity-only, adiabatic hydrodynamics, and subgrid hydrodynamics effects, all with identical initial phases -- to study the properties of halo cores, their connections across the simulations, and their relationships with galaxies. The data products used in the analysis include halo catalogs, core catalogs, core merger trees, and galaxy catalogs for the subgrid hydrodynamics simulation. 

First, we studied the statistical agreement between the core populations across the three simulations, categorizing them into central and satellite cores across several SO halo mass bins. We found, on average, 10-20\% variations in population sizes. Compared to the gravity-only simulation, the adiabatic hydrodynamics simulation has fewer cores due to gas pressure effects, while the subgrid hydrodynamics simulation has more cores due to gas cooling. Next, we compared core size distributions as a function of halo mass and halo-centric radial distance, as well as core density profiles as a function of the halo-centric radius. We again found the distributions to be in good agreement with each other, with only relatively small variations attributable to gas pressure and cooling effects. Finally, we performed a visual comparison of sample halos across the three simulations and found that the large-scale structure was qualitatively very similar, but the spatial positions of cores within halos were perturbed on small scales, mostly in non-radial directions, due to the additional physics effects included in the hydrodynamics simulations.

The above analyses show that cores are stable entities whose population statistics, size distributions, and radial distributions are largely robust against the additional effects of baryonic physics. This indicates that any method for connecting or mapping cores to galaxies within the hydrodynamics simulation can be effectively applied to cores in a gravity-only simulation to generate a realistic galaxy distribution for cosmological applications.

The next step in our analysis was to connect cores with galaxies in the subgrid hydrodynamics simulation. First, we examined the fraction of central and satellite galaxies without cores and found them to be $< 1\%$ for satellite galaxies and less than 0.1\% for central galaxies.   We further showed that the average number of cores per galaxy increases with the stellar mass of the galaxy, suggesting that the number of cores, along with their infall halo masses, could be informative for modeling stellar mass in future work.
Next, we studied the effect of applying an infall mass cut and grouping cores with different linking lengths and compared the resulting population statistics and the shapes of the halo-centric density profiles with those of galaxies. We found that an infall mass cut of $5.4\times 10^{11}$\,$h^{-1}$M$_\odot$ and a linking length of $b=0.08$\,$h^{-1}$Mpc gave the best agreement overall for these comparisons resulting in normalization differences of 10-20\% over most of the range of halo-centric distances.

With this methodology for establishing a connection between core groups and galaxies in the subgrid hydrodynamics simulation, we proceeded to apply the approach to cores in the gravity-only simulation. We compared the overall counts and the density profiles and found them to be in good agreement. The level of agreement varied depending on halo mass and halo-centric distance, suggesting potential refinements for future work. For example, a halo-mass dependent linking length could be introduced to achieve better agreement. Additionally, incorporating core velocities could refine the matching procedure. More broadly, the distributions of core velocities across the simulations warrant further investigation. 

Another possible refinement of the model would be to consider the effect of core disruption~\citep{korytov2023modeling} after infall. As with subhalos, it is possible that cores, which are initially compact objects at infall, can be pulled apart by various mass stripping effects. Further investigation of these situations in the context of cores would be an interesting research topic for the next steps in this modeling effort.
Additionally, more sophisticated core merging treatments which go beyond purely spatial criteria, are also worth investigation. 

It is important to note that the specifics of the approach developed in this analysis are tailored to the particular simulations used. We anticipate that the principle of grouping cores and selecting them via an infall mass cut will be applicable to other simulations. However, the optimal values for linking length(s) and infall mass cuts to connect cores with galaxies will vary based on the parameters of the hydrodynamics simulation, such as resolution and subgrid model implementation. In the future, we envisage that large scale gravity-only simulations will be accompanied by companion small-scale hydrodynamics simulations at the same resolution, that can be used to calibrate the galaxy-core connection.

The methodology developed in this work has strong implications for creating mock galaxy catalogs from N-body simulations. The procedure enables us to link galaxy locations and properties to those of halo cores, facilitating the creation of realistically complex galaxy catalogs while entirely bypassing the computational and resolution limits inherent in subhalo methods. 
The scalability of core-tracking algorithms renders them feasible even for large-scale simulations and the simplicity of locating cores in dense environments circumvents many problematic aspects of subhalo finding. Furthermore, the challenges of reliably determining subhalo properties for associating subhalos with galaxies are also avoided. 
Further work on the core-galaxy connection within hydrodynamics simulations, along with core correlations across simulations will enable the development of new techniques for modeling galaxies in gravity-only simulations. For example, using core merger trees combined with the core merging methods developed here could be employed to generate estimates of galaxy stellar masses and star-formation histories based on core properties. 
Although we expect galaxy stellar masses to be most strongly correlated with their core infall masses, post-infall mass estimates can be determined when necessary via modeling with SMACC~\citep{sultan2021last}.
From these basic quantities, estimates for galaxy SEDS can also be generated.

This study represents a significant advancement in mapping galaxies and their properties onto cores from gravity-only simulations. Future work aims to enhance our understanding of the dynamics of merging and the disruption of these populations, as well as to map more complex properties onto the proxy galaxies. We conclude that this method holds great promise for efficiently creating realistic large-scale mock galaxy catalogs for the modern cosmological landscape. 

\begin{acknowledgments}
\section*{Acknowledgements}
We are sincerely grateful to Salman Habib who inspired this work and provided many crucial comments throughout the development of this project. We also thank Azton Wells for producing the visualizations shown in Figs.~\ref{fig:galaxy_vis} and \ref{fig:vis_gal_core}. Finally, we thank JD Emberson for significant contributions to the hydrodynamic simulations and Esteban Rangel for leading the development of the core merger tree software. Work at Argonne National Laboratory was supported under the U.S. Department of Energy contract DE-AC02-06CH11357. This research used resources of the Argonne Leadership Computing Facility, which is supported by DOE/SC under
contract DE-AC02-06CH11357. This research also used resources of the Oak Ridge Leadership Computing Facility, which is a DOE Office of Science User Facility supported under Contract DE-AC05-00OR22725. This research was supported by the Exascale Computing Project (17-SC-20-SC), a collaborative effort of the U.S. Department of Energy Office of Science and the National Nuclear Security Administration.
E. K. was supported by the OpenUniverse effort, which is funded by NASA under JPL Contract Task 70-711320, 'Maximizing Science Exploitation of Simulated Cosmological Survey Data Across Surveys'. We thank the developers and maintainers of the following software used in our work: Python\citep{python_language}, Ipython\citep{ipython}, NumPy\citep{numpy}, Astropy\citep{astropy}, SciPy\citep{scipy}, Jupyter\citep{jupyter}, Matplotlib\citep{matplotlib}.
\end{acknowledgments}

\bibliographystyle{yahapj}
\bibliography{ref, software}

\end{document}